\documentclass[journal,draftcls,onecolumn,12pt,twoside]{IEEEtran}
\normalsize
\usepackage{amsmath}
\usepackage{amsmath,amsthm,amssymb}
\usepackage[utf8x]{inputenc}
\usepackage{color,soul}
\usepackage{amsfonts}
\usepackage{epsfig}
\usepackage{amssymb}
\usepackage{cite}
\usepackage{subfigure}
\usepackage{multirow}
\usepackage{rotating}
\usepackage{graphicx}
\usepackage{tabularx}
\usepackage{array}
\usepackage{amsfonts}
\usepackage{graphicx}  
\usepackage{graphics}  
\usepackage{epsfig}
\usepackage{tabularx}
\usepackage{amsmath}
\usepackage{amsfonts}
\usepackage{epstopdf}
\usepackage{amssymb}
\usepackage{cite}
\usepackage{subfigure}
\usepackage{multirow}
\usepackage{rotating}
\usepackage{graphicx}
\usepackage{tabularx}
\usepackage{array}
\usepackage{setspace}
\usepackage{morefloats}
\usepackage{xcolor}
\usepackage{float}
\usepackage[Euler]{upgreek}
\usepackage{mathtools}
\usepackage{algorithm} 
\usepackage[algo2e]{algorithm2e}

\usepackage[mathscr]{eucal}
\ifCLASSINFOpdf
\else
\fi

\begin{document}
%
\title{Multicasting in NOMA-Based UAV Networks: Path Design and Throughput Maximization}
%
%
%

	\author{Shima~Salarhosseini,~Mohammad~Reza~Javan,~\IEEEmembership{Senior~Member,~IEEE~}and~Ali~Nazari
		
		\thanks{ S. Salarhosseini, Mohammad. R. Javan and A. Nazari are with the Department of Electrical and Robotics Engineering, Shahrood University of Technology, Shahrood, Iran e-mail: (\{shima.salarhosseini@yahoo.com, javan@shahroodut.ac.ir and ali.nazari$\_$communication@yahoo.com\})}}

\maketitle





%

\begin{abstract}
	In this paper, we propose a new resource allocation framework for unmanned aerial vehicle (UAV) assisted multicast wireless networks in which the network users according to their request are divided into several multicast groups. We adopt power domain non-orthogonal multiple access (PD-NOMA) as the transmission technology using which the dedicated signals of multicast groups are superimposed and transmitted simultaneously as the UAV passes over the communication area for fixed and mobile users. We discuss the proposed scenarios from two perspectives, offline and online mode. In offline mode, we implement the problem for fixed and mobile users whose locations are predictable (the location of users, over the communication time, is known at the beginning of the communication time) and in online mode for mobile users whose locations are unpredictable (the location of users, over the communication time, is unknown at the beginning of the communication time). Also, we proposed a scenario in which the online model the number of mobile users can grow in each time slot. We formulate the problem of joint power allocation and UAV trajectory design as an optimization problem that is non-linear and non-convex for two proposed scenarios. To solve the problem, we adopt an alternate search method (ASM), successive convex approximation (SCA), and geometric programming (GP). Using simulation results, the performance of the proposed scheme is evaluated for different values of the network parameters.
\end{abstract}
\IEEEpeerreviewmaketitle

\section{Introduction}\label{introduction}\label{introductionmotivation}
Next generations of wireless networks would face several challenges, like high data rates and high coverage. Unmanned aerial vehicle (UAV) assisted wireless communication has attracted much attention from the research community. Especially, in areas with no communication infrastructure, they play a key role, such as providing connectivity in remote areas or areas affected by disasters or with high density of users, UAVs could be used to improve the performance of wireless networks from both the coverage as well as the data rate. There are several issues in UAV-based wireless network design that should be addressed, e.g., UAV placement, resource allocation, and trajectory design. There exist several papers which consider resource allocation and trajectory design of UAVs in wireless networks \cite{1,2,3}. In \cite{4}, the authors consider the problem of UAV placement in a wireless network such that the number of required UAVs to cover the communications area is minimized. The authors in \cite{5} consider a wireless network in which a common file should be transmitted to a set of users by a UAV. The objective is to design the trajectory of the UAV such that the transmission time is minimized. The authors in \cite{6} consider a network with multiple UAVs in which the objective is to maximize the minimum average throughput of users over the communications path. They formulate an optimization problem for joint trajectory design and transmit power allocation. They use the $\textit{\textbf{A}lternating \textbf{S}earch \textbf{M}ethod}$ (ASM) in which they solved the trajectory design problem and transmit power allocation problem in an iterative manner.  For the transmit power allocation, they use $\textit{\textbf{S}uccessive \textbf{C}onvex \textbf{A}pproximation}$ (SCA) with $\textit{\textbf{D}ifference of two \textbf{C}oncave functions}$ (D.C.) as the approximation method \cite{7}. According to the best study, the trajectory design presented in other papers hasn't considered 2D path planning of UAVs with mobile users from two perspectives of online and offline scenarios.

On the other hand, the spectrum is a scarce resource that should be used in an efficient manner. Adopting a scheme that results in high spectral efficiency is one of the concerns in wireless networks. Power domain non-orthogonal multiple access (PD-NOMA) which allows the same spectrum to be shared among more than one user has been viewed as a promising technology for improving the network spectral efficiency \cite{8,9,22}. In this technique, the users' signals are superimposed by the transmitter and the receiver applies the successive interference cancellation (SIC) method to alleviate the co-channel interference. In this regard, several works have considered PD-NOMA as the transmission technology in UAV-based networks. The authors in \cite{10} consider the downlink of a UAV-assisted wireless network in which a macrocell wants to transmit information to the network users with the help of some UAVs. In the considered model, the macrocell transmits information of users to the UAVs using the PD-NOMA scheme. Afterward, UAVs try to decode the received signal using SIC method. Next, UAVs try to transmit the decoded signals in a cooperative fashion towards the users. They aim at maximizing the total transmission rate of users by jointly determining the trajectories of UAVs and the transmit powers. To solve the optimization problem, they use ASM, SCA, and D.C., and propose an iterative solution algorithm. According to the concept of the PD-NOMA technique on the subject of SIC, the UAVs act as mobile transmitters, and mobile users act as mobile receivers; hence they need to update the SIC ordering in each time slot. 

One of the characteristics of a wireless channel is its broadcast nature in that the transmitted signal could be received by all the receivers; hence establishing multicast communications would be an interesting topic for wireless networks\cite{11}. The use of multicast communication can significantly reduce the amount of wireless network traffic, which in itself can motivate the use of this type of communication in next-generation \cite{ZhouPan}. Multicast communication can be used in practical schemes like popular programming or video conference. In \cite{LV}, the authors consider a cognitive radio network in which the primary users act as unicast users and the secondary users act as multicast users. Secondary users work as relays for the primary users. In \cite{Gau}, the authors consider a multicast network in which a base station wants to transmit some common information towards the network users forming several multicast groups. The intended signals are transmitted simultaneously towards users using the PD-NOMA technique. The objective is to determine the user association and transmit power allocation such that the total transmission rate of the network users is maximized. The authors in \cite{Nasir} consider a network in which a UAV wants to transmit common files to ground users with multicasting. The objective is to design the UAV trajectory to minimize the total transmission time such that the ground users receive their intended information successfully with high probability.  In \cite{15}, the authors consider the problem of joint transmit power allocation, UAV's altitude determination, and antenna beamwidth design in a non-orthogonal multiple access (NOMA) based wireless network. Their objective is to maximize the minimum rate of users under transmit power and UAV altitude constraints. UAVs have limitations in energy consumption, flying time, and serving the agents. On the other hand, multicast-capable networks allow one or more sources to efficiently send data to a group of recipients that requested or needed the same type of information. Hence, a combination of UAV in multicast-capable networks causes energy efficiency improvement.  

\subsection{Motivation and Contributions}\label{subsec1.1}
In this paper, we propose a resource allocation framework for unmanned aerial vehicle (UAV) assisted multicast wireless networks in which the network users according to their requests are divided into several multicast groups. We adopt power domain non-orthogonal multiple access (PD-NOMA) as the transmission technology using which the signals of multicast groups are superimposed and transmitted simultaneously as the UAV passes over the communication area.

The main contributions of this paper can be summarized
as follows:
\begin{itemize}
	\item
	In this paper, we formulate the resource allocation problem into an optimization problem to jointly determine the trajectory of the UAV and the transmit power of the UAV over its trajectory, in which the UAV serves several multicast groups. The objective is to maximize the total transmission rate of multicast groups over the communications time while a minimum transmission rate should be guaranteed for each multicast group, transmit power constraint should be satisfied, and the maximum speed limitation for the UAV should be taken into account.
	\item
	We implement two online and offline models for the proposed system, which has not been seen in any similar scenario. In offline mode, we have designed two approaches: the fixed users and also the mobile users with predictable movement (the location of users, over the communication interval, are known at the beginning of the communication). In online mode, we deal with mobile users with unpredictable movements. In addition, we have made it possible for all users to be able to move in different geographical directions and at different speeds. 
	\item
	The resulting optimization problem is non-linear and non-convex obtaining whose solution is very hard. To tackle this, we adopt the ASM method where the main problem is decomposed into trajectory design problem and transmit power determination problem. For the trajectory design, we use $\textit{\textbf{G}eometric \textbf{P}rograming}$ (GP) and for the transmit power design, we use successive convex approximation with D.C. method as the approximation technique.
	\item
	In the simulation part, we investigate the different parameters, such as the effect of time on the total rate and trajectory design of UAV, trajectory design of mobile user with predictable location and mobile user with the unpredictable location that because UAV should arrive at end point (destination point) we propose new constraint that satisfies the successful path design for mobile user with the unpredictable location.
	\item
	As another innovation implemented in this article, we can refer to the scenario of changing the number of users. We consider the case that the number of users in each multicast group can be changed in each time slot. Variation of the number of users besides mobility of them in the online scenario can bring a real and practical model.
\end{itemize}

\subsection{Paper Organisation}\label{subsec1.2}

The rest of this paper is organized as follows. In Section
\ref{systemmodelproblemformulation}, the system model and the problem formulation are discussed in detail. In Section \ref{PROPOSED ALGORITHM}, the non-convex optimization problem is solved 
. We consider the case of mobile users in Section \ref{Mobile User Location}. The study of computational complexity and convergence of the proposed schemes is provided in Section \ref{COMPUTATIONAL COMPLEXITY AND CONVERGENCE ANALYSIS}.
In Section \ref{sec:simulation}, we provide simulations verifying the performance of the developed algorithms. Finally, the paper is concluded in Section \ref{sec:conclusion}. 

\begin{figure}[t]
	\begin{center}
		\includegraphics[width=10 cm]{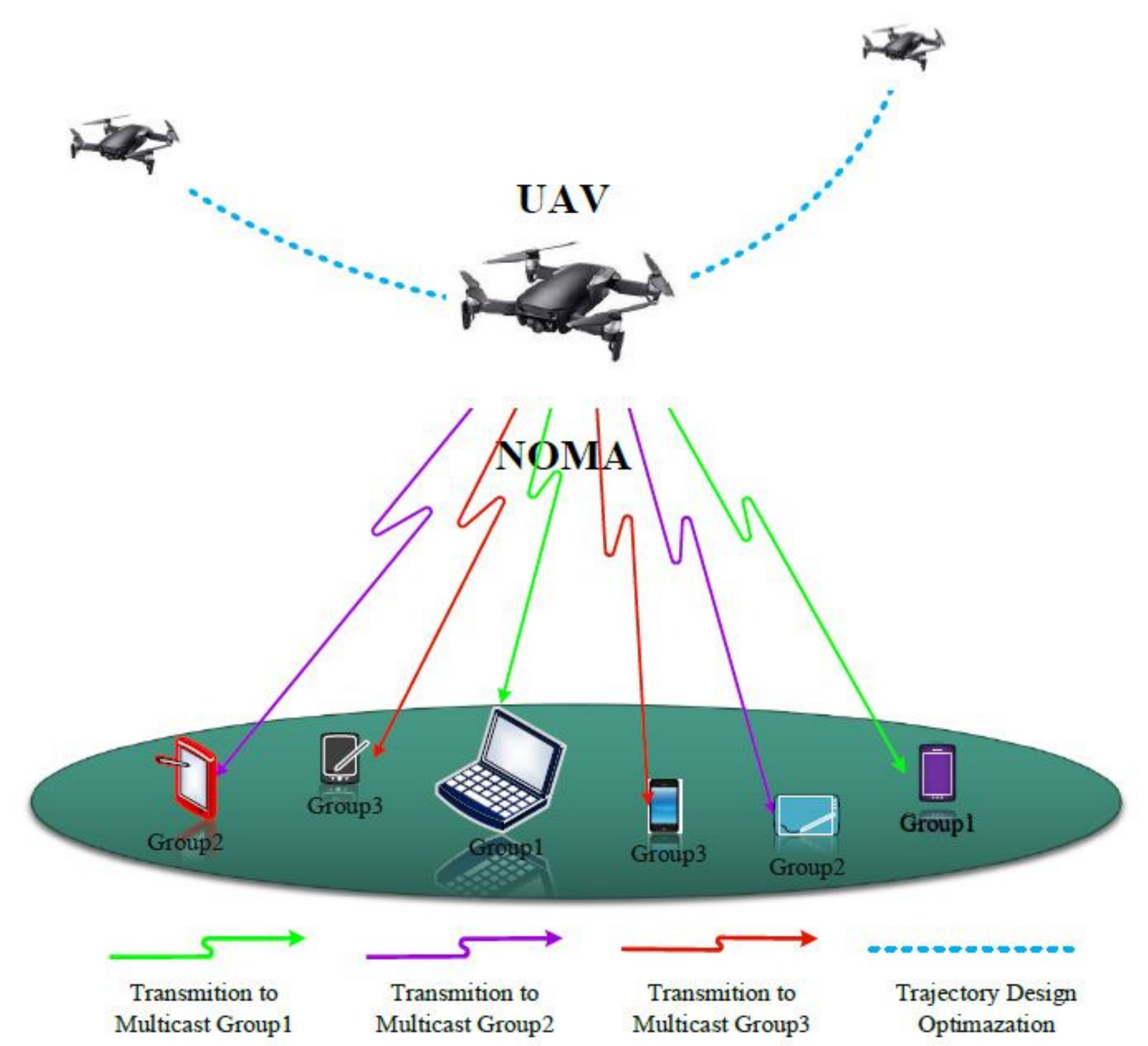} \vspace{-0.6cm}
		\caption{A UAV enabled wireless network with multicast communication.}\label{fig:sim_conf}\vspace{-0.1cm}
	\end{center}\vspace{-.3cm}
\end{figure} 
\vspace{-0.25cm}

\section{System Model and Problem formulation}\label{systemmodelproblemformulation}

\subsection{System Model}\label{sysmodel}

We consider a communication scenario in which a UAV is responsible for information transmission towards the network's users and UAV flies over the network area for a period of time over a communication path, and according to the concept of PD-NOMA, UAV sends the proportional message towards the network users of each multicast group, simultaneously.
We assume the users according to their request are partitioned into $G$ multicast groups.
The set of users in each group $g\in \mathcal{G}=\{1,\cdots,G\}$ is denoted by $u_g \in \mathcal{U}_g=\{1,\cdots, U_g\}$, and the set of total users is denoted by $ \mathcal{U}=\bigcup\limits_{g \in \mathcal{G}} \mathcal{U}_g $ with cardinality $U=\sum \limits_{g \in \mathcal{G}} U_g$.
The system model considered in the paper is shown in Fig. \ref{fig:sim_conf}. We also assume that the UAV uses the PD-NOMA as the transmission technology where the information signals of all the multicast groups ($G$ information signals) are superimposed and the resulting signal is transmitted towards the users. Each user applies the SIC technique to extract the transmitted information. To this end, we should determine the SIC ordering of users such that each user knows that the information of which users should be detected and canceled from the received signal. Since we consider users are located in multicast groups, UAV transmits signals as multicasting signals toward users. Hence, all the users in the same group perform the SIC in the same manner and we should only determine the order of multicast groups.
In other words, the users of each group detect and cancel the same signals (users according to their request are partitioned into the multicast groups), i.e., those from the groups which are specified by the SIC ordering process. The users are assumed to be randomly located over the network coverage area and to be fixed over the considered time interval. We denote the location of user $u_g$ by $\mathbf{r}_{g,u_g}=[\ x_{g,u_g},y_{g,u_g}]^\text{T}\in \mathbb{R}^{2 \times 1}, $ where $[\cdot]^\text{T}$ denotes the vector transpose. The UAV is assumed to be mobile and could move over the coverage area at the maximum speed of $V_\text{max}$ meter per second (m/s) with the fixed altitude of $H$ meters. We study the communications of the considered network over a time interval of $T$ seconds. During this time, the UAV starts its trajectory from a start point in the network towards its end point while it transmits information to the network users. As is common in the literature \cite{20}, we consider the discrete model in which the total time is divided into a number of time slots, e.g., $N$, each of which is so much small such that we can assume that the channel gain from UAV to the network users are fixed. We call the positions of the UAV at each of these discrete-time instances the breaking points. Therefore, we have $N+1$ breaking points whose set is denoted by $\mathcal{N}=\{0,1,\cdots,N\}$ with $n=0$ for the start point and $n=N$ for the end point.

The positions of the UAV at the beginning and the end of the time slot $n$ are, respectively, denoted by $\mathbf{q}[\mathnormal{n-1}]=[x_q[\mathnormal{n-1}],y_q[\mathnormal{n-1}]]^T$ and $\mathbf{q}[\mathnormal{n}]=[x_q[\mathnormal{n}],y_q[\mathnormal{n}]]^T$, which are respectively, breaking points $n-1$ and $n$. Over each time slot, the UAV transmits information towards the users in all of the $G$ multicast groups. We define  $\mathbf{Q}=\{\mathbf{q}[n], \forall \mathnormal{n} \in \mathcal{N}\}$ for the trajectory design, we restrict the distance between two consecutive breaking points by imposing the following constraint:
\begin{align}\label{speed}
{\Vert \mathbf{q}[n] - \mathbf{q}[n-1] \Vert}^2 \leqslant \mathnormal{S}^2_\text{max},~~n=1,...,N,
\end{align}
where $\mathnormal{S}_\text{max} = \frac{\mathnormal{V}_\text{max}T}{N}$, and we emphasis that the start point, i.e., $\mathbf{q}[0]$ and the end point, i.e., $\mathbf{q}[N]$, are known.

For the channel gain, we only consider pathloss which means that the channel from UAV to user $u_g$ is assumed to follow the free-space pathloss model. We assume that the channel gain from UAV to user $u_g$ time slot $n$ is given by

\begin{align}\label{pathlossmodel}
h_{g,u_g}[n]= \frac{\mu_0}{H^2+{\Vert \mathbf{q}[n] - \mathbf{r}_{g,u_g} \Vert}^2},
\end{align}
where $\mu_0$ is the pathloss at the reference distance $\hat{d} = 1$ meter and  we note that $\sqrt{H^2+{\Vert \mathbf{q}[n] - \mathbf{r}_{g,u_g} \Vert}^2}$ is the distance from UAV to user $u_g$ in multicast group $g$ when UAV is at breaking point $n$.

We define the channel gain representative of users $u_g$ in group $g$ in time slot $n$ by
\begin{align}\label{channelpowergain}
\hat{h}_{g}[n]=\min_{u_g \in \mathcal{U}_g}h_{g,u_g}[n],
\end{align}
which is the minimum channel gain of users $u_g$ in the multicast group $g$ at breaking point $n$.

Let $p_g[n]$ be the transmit power of UAV used for information transmission to multicast group $g$ over time slot $n$. We define $\mathbf{p}[n]=[p_1[n],p_2[n],\cdots,p_G[n]]^T$, $\mathbf{p}_g=[p_g[1],p_g[2],\cdots,p_g[N]]^T$, $\mathbf{p}=[\mathbf{p}_1,\mathbf{p}_2,\cdots,\mathbf{p}_G]^T$. The achievable multicast capacity of users $u_g$ in group $g$ in time slot $n$ is given by
\begin{align}\label{achievable multicast capacity}
C_g[n]=\ln({1+\frac{p_g[n] \hat{h}_g[n]}{I_{g}[n]+\sigma^2}}),~~\forall g \in \mathcal{G},~~n=1,\cdots,N,
\end{align}
where $\sigma^2$ is the additive white Gaussian noise (AWGN) power at the receiver which is assumed to be the same for all the receivers and ${I}_g[n]$ in \eqref{interference} is the interference caused by NOMA transmission as given by 
\begin{align}\label{interference}
I_g[n]=\hat{h}_g[n] \sum\limits_{j \in \mathcal{O}_g[n]} p_j[n], 
\end{align}
where
\begin{align}\label{ordering}
\mathcal{O}_g[n]=\{ 
j \in \mathcal{G} \vert~ \hat{h}_j[n] > \hat{h}_g[n]
\},
\end{align}
which represents the set of multicast groups whose signals could not be canceled by users $u_g$ in the multicast group $g$. In other words, this determined the SIC ordering in NOMA-based communications.
\subsection{Problem Formulation}\label{Problem Formulation}
Now we provide our proposed optimization problem. The objective is to maximize the sum multicast capacities of multicast groups over the communication time interval. For the quality of service (QoS) purpose, we require that a minimum multicast capacity for each multicast group should be satisfied. We also have transmit power constraint for the UAV as well as the maximum distance limitation between two consecutive breaking points. Therefore, our proposed optimization problem is given by
\begin{subequations}\label{optimizationproblemmain}
	\begin{align}
	\max\limits_{\mathbf{p},\mathbf{Q}}&\sum\limits_{n =1}^{N}\sum\limits_{g \in \mathcal{G}}C_g[n]
	\\
	\label{optimizationproblemmatext1}
	\text{s.t.}~~&{\Vert \mathbf{q}[n] - \mathbf{q}[n-1] \Vert}^2 \leqslant S^2_\text{max},~~n=1,\dots,N,
	\\
	\label{optimizationproblemmasum1}
	&
	C_g[n] \geqslant C_g^{\text{rsv}}[n],~~\forall g \in \mathcal{G},~~n=1,\dots,N,
	\\
	\label{optimizationproblemmatwosum}
	&
	\sum\limits_{g \in \mathcal{G}}p_{g}[n]\leqslant P_{\text{max}},~~~~n=1,\dots,N,
	\end{align}
\end{subequations}
where \eqref{optimizationproblemmatext1} ensures that the maximum distance between two consecutive points is limited, \eqref{optimizationproblemmasum1} is the minimum multicast capacity constraint for each multicast group, and \eqref{optimizationproblemmatwosum} is the transmit power constraint. Note that in constraint \eqref{optimizationproblemmasum1}, $C_g^{\text{rsv}}[n]$ is the minimum required data rate of group $g$ in time slot $n$ and $P_{\text{max}}$ is the maximum transmit power for each time slot in constraint \eqref{optimizationproblemmatwosum}.

\section{PROPOSED ALGORITHM}\label{PROPOSED ALGORITHM}
We note that the optimization problem \eqref{optimizationproblemmain} is a non-linear and non-convex problem that is generally hard to solve. To tackle this problem, we devise an alternating method in which we first solve the problem for some of the optimization variables while the others are considered fixed, and then given the solution of this step, we solve the optimization problem for the remaining optimization variables. Therefore, we divide our main problem into two optimization problems: UAV trajectory design in which the transmit power is assumed to be fixed, and the transmit power allocation in which the breaking points are fixed.


\subsection{UAV Trajectory Optimization with Fixed power Allocation}\label{UAV Trajectory Optimization with Fixed Power Allocation}
To solve the optimization problem and find the trajectory of the UAV with fixed transmit powers, we rely on the successive convex approximation method and geometric programming (GP). This means that we solve a sequence of convex approximated problems whose convergence to a sub-optimal solution is guaranteed.

For the fixed transmit power variables, the optimization problem \eqref{optimizationproblemmain} is transformed into the following one:
\begin{subequations}\label{optmiationproblemtrajectorymain}
	\begin{align}
	\label{00}
	\max\limits_{\mathbf{Q},\mathbf{\varPsi}}&\sum\limits_{n =1}^{N}\sum\limits_{g \in \mathcal{G}}\ln({1+\frac{p_g[n] \hat{h}_g[n]}{\varPsi_g[n]}})
	\\
	\label{01}
	\text{s.t.}~~&{\Vert \mathbf{q}[n] - \mathbf{q}[n-1] \Vert}^2 \leqslant S^2_\text{max},~~n=1,\cdots,N,
	\\
	\label{02}
	&
	\ln(1+\frac{p_g[n] \hat{h}_g[n]}{\varPsi_g[n]}) \geqslant C_g^{\text{rsv}}[n],~~\forall g \in \mathcal{G},~~n=1,\cdots,N-1,\\
	\label{03}
	&I_g[n]+\sigma^2 \leqslant \varPsi_g[n],~~\forall g \in \mathcal{G},~~~~n=1,\cdots,N-1,
	\end{align}
\end{subequations}
where $\mathbf{\varPsi}[n]=[\varPsi_1[n],\varPsi_2[n],\cdots,\varPsi_G[n]]^T$, $\mathbf{\varPsi}_g=[\varPsi_g[1],\varPsi_g[2],\cdots,$ $\varPsi_g[N]]^T$, and $\mathbf{\varPsi}=\text{vec}[\mathbf{\varPsi}_1,\mathbf{\varPsi}_2,\cdots,\mathbf{\varPsi}_G]^T$, and we note that a new variable $\varPsi_g[n]$ is defined and a new constraint (6d) is added.

Note that the optimization problem \eqref{optmiationproblemtrajectorymain} could not be directly transformed into the GP form and some (convex) approximations are needed to make the constraint transformable into the GP form. This means that we will devise an SCA-based algorithm to find the trajectory of UAV in an iterative manner. First, the constraint \eqref{01} is rewritten as follows:
\begin{align}\label{gpconst}
&(x_q[n]-x_q[n-1])^2+(y_q[n]-y_q[n-1])^2 \leqslant S^2_\text{max}.
\end{align}
Next, we expand the terms in \eqref{gpconst} as follows:
\begin{align}\label{expterms}
{x_q[n]}^2+{x_q[n-1]}^2+{y_q[n]}^2+{y_q[n-1]}^2\leqslant
\nonumber
\\ 2x_q[n]x_q[n-1]+2y_q[n]y_q[n-1]+S^2_\text{max}.
\end{align}
We divide both sides of \eqref{expterms} by the $2x_q[n]x_q[n-1]\\+2y_q[n]y_q[n-1]+S^2_\text{max}$ to obtain the following:
\begin{align}\label{divdterm}
&\frac{{x_q[n]}^2+{x_q[n-1]}^2+{y_q[n]}^2+{y_q[n-1]}^2}{2x_q[n]x_q[n-1]+2y_q[n]y_q[n-1]+S^2_\text{max}} \leqslant 1.
\end{align}
Next, we approximate the denominator of \eqref{divdterm} as a posynomial \cite{16} to obtain a convex form for the constraint \eqref{01} as follows:
\begin{align}\label{cvxterms}
&\Big({x_q[n]}^2+{x_q[n-1]}^2+{y_q[n]}^2+{y_q[n-1]}^2\Big)\nonumber \\ \times &\Big(\frac{2 x_q[n] x_q[n-1]}{\alpha^{t_1}[n]}\Big)^{-\alpha^{t_1}[n]}\times \Big(\frac{2 y_q[n] y_q[n-1]}{\beta^{t_1}[n]}\Big)^{-\beta^{t_1}[n]}
\nonumber \\ 
\times &\Big(\frac{S^2_\text{max}}{\gamma^{t_1}[n]}\Big)^{-\gamma^{t_1}[n]} \leqslant 1, 
\end{align}
where
\begin{align}
\label{alpha}
&\alpha^{t_1}[n]=\frac{2 x_q^{t_1-1}[n] x_q^{t_1-1}[n-1]}{2x_q^{t_1-1}[n]x_q^{t_1-1}[n-1]+2y_q^{t_1-1}[n]y_q^{t_1-1}[n-1]+S^2_\text{max}},
\\
\label{beta}
&\beta^{t_1}[n]=\frac{2 y_q^{t_1-1}[n] y_q^{t_1-1}[n-1]}{2x_q^{t_1-1}[n]x_q^{t_1-1}[n-1]+2y_q^{t_1-1}[n]y_q^{t_1-1}[n-1]+S^2_\text{max}},
\\
\label{gammahat}
&\gamma^{t_1}[n]=\frac{S^2_{max}}{2x_q^{t_1-1}[n]x_q^{t_1-1}[n-1]+2y_q^{t_1-1}[n]y_q^{t_1-1}[n-1]+S^2_\text{max}}.
\end{align}

To obtain a convex approximation for the objective function \eqref{00}, we do as follows:
\begin{subequations}\label{objfuncsub}
	\begin{align}
	\label{10}
	&\max\limits_{\substack{x_q[n], y_q[n]\\ ,L_g[n] , \varPsi_g[n]}}~
	\sum\limits_{n =1}^{N}\sum\limits_{g \in \mathcal{G} }\ln\Big(1+\frac{p^{t-1}_g[n]\mu_0}{L_g[n] \varPsi_g[n]}\Big)
	\\
	\label{11}
	=&\max\limits_{\substack{x_q[n], y_q[n]\\ ,L_g[n] , \varPsi_g[n]}}~\ln\Big(\prod\limits_{n =1}^{N}\prod\limits_{g \in \mathcal{G}}\frac{L_g[n] \varPsi_g[n]+p^{t-1}_g[n]\mu_0}{L_g[n] \varPsi_g[n]}\Big)
	\\
	\label{12}
	=&\min\limits_{\substack{x_q[n], y_q[n]\\ ,L_g[n] , \varPsi_g[n]}}~\ln\Big(\prod\limits_{n =1}^{N}\prod\limits_{g \in \mathcal{G}}\frac{L_g[n] \varPsi_g[n]}{L_g[n] \varPsi_g[n]+p^{t-1}_g[n]\mu_0}\Big)
	\\
	\label{13}
	\doteq&\min\limits_{\substack{x_q[n], y_q[n]\\ ,L_g[n], \varPsi_g[n]}}~\prod\limits_{n =1}^{N}\prod\limits_{g \in \mathcal{G}}\frac{L_g[n] \varPsi_g[n]}{L_g[n] \varPsi_g[n]+p^{t-1}_g[n]\mu_0},
	\end{align}
\end{subequations}
where $\doteq$ in \eqref{13} denotes the equivalency between terms and $t$ is the iteration number of the ASM algorithm.
We note that a new optimization variable, i.e., $L_g[n]$, is introduced which together with \eqref{channelpowergain}, we have:
\begin{align}\label{optvar}
\max_{u_g\in\mathcal{U}_g}(H^2+{\Vert \mathbf{q}[n] - \mathbf{r}_{g,u_g} \Vert}^2)&=L_g[n],\\~~\forall g \in \mathcal{G},&~~~~n=1,\cdots,N-1.\nonumber
\end{align}
Therefore, the following constraints should be added to the optimization problem:
\begin{align}\label{focons}
H^2+{\Vert \mathbf{q}[n] - \mathbf{r}_{g,u_g} \Vert}^2\leqslant &L_g[n],\\&\forall u_g \in \mathcal{U}_g,~~\forall g \in \mathcal{G},
~~n=1,\cdots,N-1\nonumber.
\end{align}
Similar to \eqref{gpconst}, one can obtain the following approximation for \eqref{focons}:
\begin{align}\label{appcons}
&\Big({x_q[n]}^2+{x_{g,u_g}}^2+{y_q[n]}^2+{y_{g,u_g}}^2+H^2\Big) \nonumber \\
\times &\Big(\frac{2x_q[n]x_{g,u_g}}{\eta_{g,u_g}^{t_1}[n]}\Big)^{-\eta_{g,u_g}^{t_1}[n]}
\times \Big(\frac{2y_q[n]y_{g,u_g}}{\kappa_{g,u_g}^{t_1}[n]}\Big)^{-\kappa_{g,u_g}^{t_1}[n]} \nonumber
\\
\times &\Big(\frac{ L_g[n]}{\vartheta_{g,u_g}^{t_1}[n]}\Big)^{-\vartheta_{g.u_g}^{t_1}[n]} \leqslant 1,
\end{align}
where
\begin{align}
\label{eta}
\eta_{g,u_g}^{t_1}[n]=\frac{2x_q^{t_1-1}[n]x_{g,u_g}}{2x_q^{t_1-1}[n]x_{g,u_g}+2y_q^{t_1-1}[n]y_{g,u_g}+L_g^{t_{1}-1}[n]},
\\
\label{kappa}
\kappa_{g,u_g}^{t_1}[n]=\frac{2y_q^{t_1-1}[n]y_{g,u_g}}{2x_q^{t_1-1}[n]x_{g,u_g}+2y_q^{t_1-1}[n]y_{g,u_g}+L_g^{t_{1}-1}[n]},
\\
\label{vartheta}
\vartheta_{g,u_g}^{t_1}[n]=\frac{L_g^{t_{1}-1}[n]}{2x_q^{t_1-1}[n]x_{g,u_g}+2y_q^{t_1-1}[n]y_{g,u_g}+L_g^{t_{1}-1}[n]}.
\end{align}

Next, we approximate the denominator of each term in \eqref{objfuncsub} by a monomial to obtain $
\prod\limits_{n =1}^{N}\prod\limits_{g \in \mathcal{G}}\Gamma_g[n]$ where
\begin{align}
\label{gamma}
\Gamma_g[n]=&\big(L_g[n] \varPsi_g[n]\big)\times \Big(\frac{L_g[n] \varPsi_g[n]}{\nu_g^{t_1}[n] }\Big)^{-\nu_g^{t_1}[n]} \nonumber \\ \times &\Big(\frac{p^{t-1}_g[n]\mu_0}{\xi_g^{t_1}[n]}\Big)^{-\xi_g^{t_1}[n]},\\
\label{nu}
&\nu_g^{t_1}[n]=\frac{L_g^{t_1-1}[n] \varPsi_g^{t_1-1}[n]}{L_g^{t_1-1}[n] \varPsi_g^{t_1-1}[n]+p^{t-1}_g[n]\mu_0},
\\
\label{xi}
&\xi_g^{t_1}[n]=\frac{p^{t-1}_g[n]\mu_0}{L_g^{t_1-1}[n] \varPsi_g^{t_1-1}[n]+p^{t-1}_g[n]\mu_0}.
\end{align}

Similarly, the constraint \eqref{02} could be written as follows:
\begin{align}\label{similarcons}
\Gamma_g[n] \leqslant \exp(-C_g^{\text{rsv}}[n]),~~\forall g \in \mathcal{G},~~n=1,\cdots,N-1,
\end{align}
where $\exp(\cdot)$ is an exponential function.

Finally, we can write \eqref{03} as follows:
\begin{align}	\label{03new}
\Big(\hat{h}_g[n] \sum\limits_{j \in \mathcal{O}_g[n]} p^{t-1}_j[n]+\sigma^2\Big)&\times\Big(\varPsi_g[n]\Big)^{-1} \leqslant 1 ,\\&~~\forall g \in \mathcal{G}, ~~n=1,\cdots,N-1\nonumber.
\end{align}

Therefore problem \eqref{optmiationproblemtrajectorymain} could be approximated as a general GP form as follow:
\begin{subequations}\label{gptransform}
	\begin{align}\label{monomial2}
	&\min\limits_{\substack{x_q[n], y_q[n]\\ ,L_g[n] , \varPsi_g[n]}}~\prod\limits_{n =1}^{N}\prod\limits_{g \in \mathcal{G}}\Gamma_g[n],
	\\
	&~~\text{s.t.}~\eqref{cvxterms},~\eqref{appcons},~\eqref{similarcons},~\eqref{03new},
	\end{align}
\end{subequations}
which can be solved by existing software toolboxes like CVX \cite{18}.

\subsection{Power Allocation Optimization with Fixed UAV Trajectory}\label{Power Allocation Optimization with Fixed UAV Trajectory}

To solve the transmit power optimization problem given the trajectory of the UAV, we use a successive convex approximation method based on the D.C. method. With the trajectory of the UAV obtained in the previous subsection, we solve the following problem to obtain the transmit powers:
\begin{subequations}\label{optmiationproblempowermain}
	\begin{align}
	\label{maxprob}\max\limits_{\mathbf{p}}& \sum\limits_{n =1}^{N}\sum\limits_{g \in \mathcal{G}}\ln(1+\frac{p_g[n] \hat{h}^{t}_g[n]}{I_g[n]+\sigma^2})
	\\
	\label{maxprobst}\text{s.t.}~~&
	\ln(1+\frac{p_g[n] \hat{h}^{t}_g[n]}{I_g[n]+\sigma^2}) \geqslant C_g^{\text{rsv}}[n],~~\forall g \in \mathcal{G},~~n=1,\cdots,N,
	\\
	\label{maxprobsum}&
	\sum\limits_{g \in \mathcal{G}}p_{g}[n] \leqslant P_{\text{max}},~~n=1,\cdots,N,
	\end{align}
\end{subequations}
where $\hat{h}^{t}_g[n] = \min_{u_g \in \mathcal{U}_g}\frac{\mu_0}{H^2+{\Vert \mathbf{q}^{t}[n] - \mathbf{r}_{g,u_g} \Vert}^2}.$
\\
We note that the optimization problem \eqref{optmiationproblempowermain} is non-convex due to the objective function and the constraint \eqref{maxprobst} which should be approximated by convex functions. We rewrite the objective function \eqref{maxprob} as follows:
\begin{align}\label{convexfun}
\sum\limits_{n =1}^{N}\sum\limits_{g \in \mathcal{G}}\ln(1+\frac{p_g[n] \hat{h}^{t}_g[n]}{I_g[n]+\sigma^2})=\sum\limits_{n =1}^{N}\sum\limits_{g \in \mathcal{G}} (F_g[n] - G_g[n]),
\end{align}
where 
\begin{align}\label{ffunc}
&F_g[n]=\ln \Big(p_g[n] \hat{h}^{t}_g[n] + \hat{h}^{t}_g[n] \times\sum\limits_{j \in \mathcal{O}_g[n]}p_j[n] + \sigma^2\Big),
\\
\label{gfunc}
&G_g[n]=\ln \Big(\hat{h}^{t}_g[n] \times\sum\limits_{j \in \mathcal{O}_g[n]}p_j[n] + \sigma^2\Big),
\end{align}
are concave functions. Therefore, we approximation functions \eqref{gfunc} as follows:
\begin{align}\label{concavefunc}
\hat{G}_{g}[n] &\approx \ln(\hat{h}^{t}_g[n]\times\sum\limits_{j \in \mathcal{O}_i[n]}p_j^{t_2-1}[n]+\sigma^2) \nonumber \\&+\sum\limits_{j \in \mathcal{O}_i[n]}\nabla G_j[n](p_j[n]-p_j^{t_2-1}[n]),
\end{align}
where
\begin{align}\label{nabla}
\nabla G_j[n]=\frac{\hat{h}^{t}_g[n]}{\hat{h}^{t}_g[n] \times \sum\limits_{j \in \mathcal{O}_i[n]}p_j^{t_2-1}[n]+\sigma^2},
\end{align}
where $t_2$ is the iteration index of the SCA-based solution algorithm to find the transmit power of UAV. Note that using the above results, a similar approximation could be obtained for \eqref{maxprobst}. Therefore, the convex approximation of the problem \eqref{optmiationproblempowermain} is given by
\begin{subequations}\label{optmiationproblempowermaingp}
	\begin{align}
	\label{maxprobgp}\max\limits_{\mathbf{p}}& \sum\limits_{n =1}^{N}\sum\limits_{g \in \mathcal{G}}(F_g[n]-\hat{G}_g[n])
	\\
	\label{maxprobstgp}\text{s.t.}~~&
	(F_g[n]-\hat{G}_g[n])\geqslant C_g^{\text{rsv}}[n],~~\forall g \in \mathcal{G},~~n=1,\cdots,N,	
	\\ 
	\label{maxprobsumgp}&
	\sum\limits_{g \in \mathcal{G}}p_{g}[n] \leqslant P_{\text{max}},~~n=1,\cdots,N,
	\end{align}
\end{subequations}
which can be efficiently solved by the existing solvers like CVX \cite{18}.

A high level view of the structure of the proposed solution scheme for the optimization problem \eqref{optimizationproblemmain} is shown as a flow chart and the detailed solution is given in Algorithm 1.
\begin{figure}[]
	\begin{center}
		\includegraphics[width=9 cm , height=9 cm]{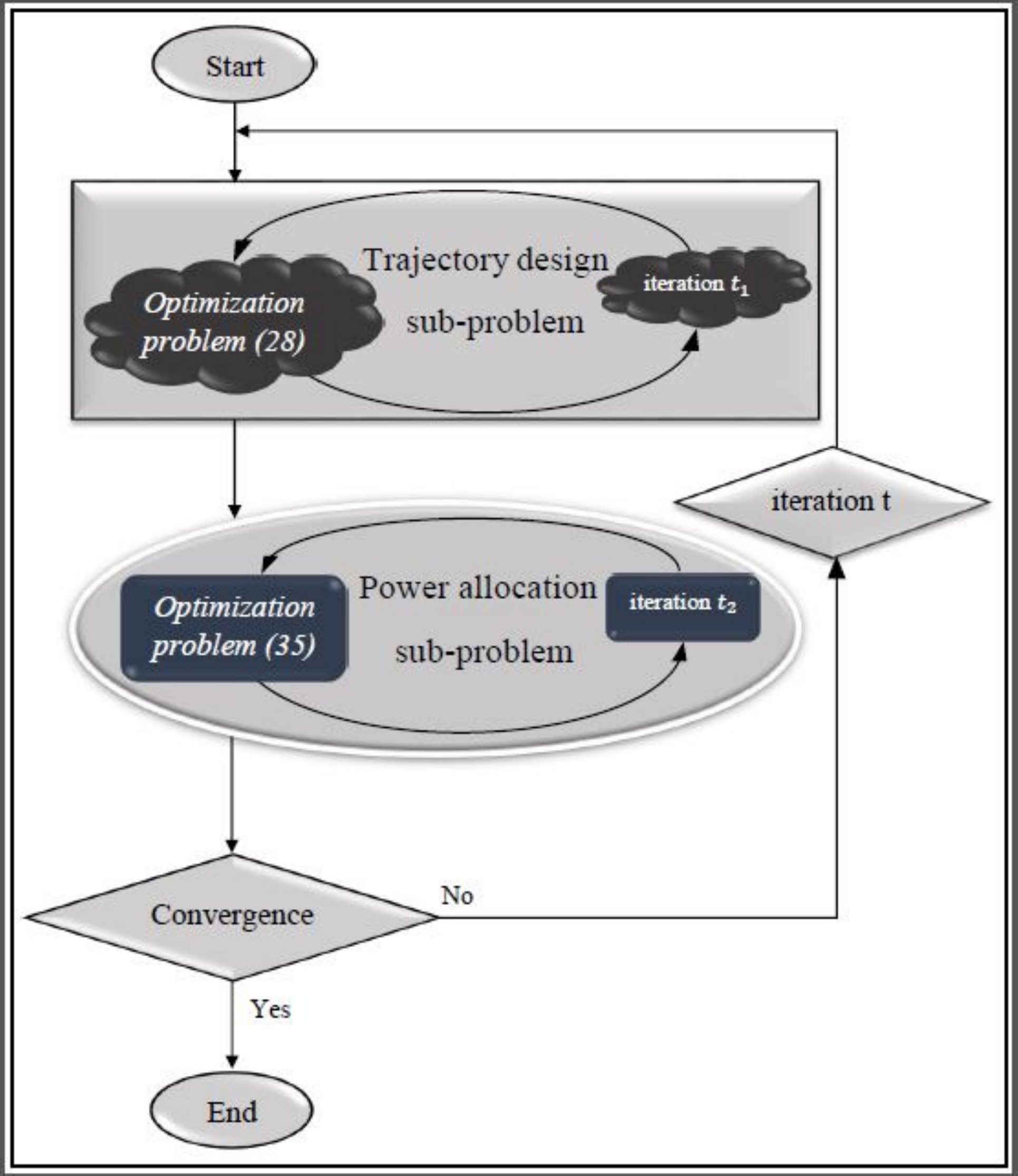} \vspace{-0.25cm}
		\caption{The flow chart for the main optimization problem \eqref{optimizationproblemmain}: See Algorithm 1 for more details}\label{fig:flowchart} \vspace{-0.1cm}
	\end{center}\vspace{0.3cm}
\end{figure} 

\begin{center}\label{Algtrajectory}
	\begin{footnotesize}\label{Algtrajectory}
		\begin{tabular}{l}
			\hline
			\textbf{Algorithm 1: Solution algorithm for the main optimization problem \ref{optimizationproblemmain}}\\
			\hline
			1.
			Set $t=0$ and initialize $\mathbf{p}^t$ and $\mathbf{Q}^t$.
			\\
			2.
			\textbf{Repeat}
			\\\vspace{0.05cm}\textbf{SCA-GP-based procedure for solving Problem \eqref{gptransform}}
			\\\vspace{0.05cm}
			2.1.1.
			Set $t_1=0$ and initialize $\mathbf{Q}^{t_1}=\mathbf{Q}^t,~L_g^{t_1}[n],~\varPsi_g^{t_1}[n],~\forall n,~\forall g.$
			\\
			2.1.2.
			\textbf{Repeat}
			\\
			2.1.3.
			Comput $\alpha^{t_1+1}[n],~\beta^{t_1+1}[n],~\gamma^{t_1+1}[n]$~from \eqref{alpha}~ to ~\eqref{gammahat}.
			\\
			2.1.4.
			Comput ~$\eta^{t_1+1}[n],~\kappa^{t_1+1}[n],~\vartheta^{t_1+1}[n]$ from \eqref{eta} to \eqref{vartheta}.
			\\
			2.1.5.
			Comput ~$\nu^{t_1+1}[n],~\xi^{t_1+1}[n]$~from ~\eqref{nu},~\eqref{xi}.
			\\
			2.1.6.
			Solve the problem \eqref{gptransform} for to find ~$\mathbf{Q}^{{t}_1+1}$, $L^{{t}_1+1}$ and $\varPsi^{{t}_1+1}$
			\\
			~~~~~~~~using geometric programing in CVX \cite{18}.
			\\
			2.1.7.
			Set~$t_1=t_1+1$.
			\\
			2.1.8.
			\textbf{Until}~~$\Vert {\mathbf{Q}^{t_1}} -{\mathbf{Q}^{t_1-1}} \Vert \leqslant \varepsilon_1.$
			\\
			2.1.9.
			$\mathbf{Q}^{t+1}=\mathbf{Q}^{t_1}$.
			\\	\vspace{0.05cm}\textbf{SCA-D.C.-based procedure for solving Problem \eqref{optmiationproblempowermaingp}}
			\\\vspace{0.05cm}
			2.2.1.
			Set $t_2=0$ and initialize $\mathbf{p}^{t_2}=\mathbf{p}^t.$
			\\
			2.2.2.
			$\textbf{Repeat}$
			\\
			2.2.3.
			Solve the problem \eqref{optmiationproblempowermaingp} for to find $\mathbf{p}^{{t_2}+1}$ using CVX \cite{18}.
			\\
			2.2.4.
			Set $t_2=t_2+1$.
			\\
			2.2.5.
			\textbf{Until}~~$\Vert {\mathbf{p}^{t_2}} -{\mathbf{p}^{t_2-1}} \Vert \leqslant \varepsilon_2.$
			\\
			2.2.6.
			$\mathbf{p}^{t+1}=\mathbf{p}^{t_2}$.
			\\\vspace{0.05cm}
			3.
			Set $t=t+1$.
			\\
			4.
			\textbf{Until}~~$\Vert {\mathbf{Q}^{t}} -{\mathbf{Q}^{t-1}} \Vert \leqslant \varepsilon_1$ and 
			$\Vert {\mathbf{p}^{t}} -{\mathbf{p}^{t-1}} \Vert \leqslant \varepsilon_2$.
			\\
			\hline\vspace{0.05cm}
		\end{tabular}\vspace{-0.5cm}
	\end{footnotesize}\label{Algtrajectory}
\end{center}

	\section{The Case of Mobile Users}\label{Mobile User Location}
	
\begin{figure}[]
		\begin{center}
			\includegraphics[width=9 cm , height=5.9 cm]{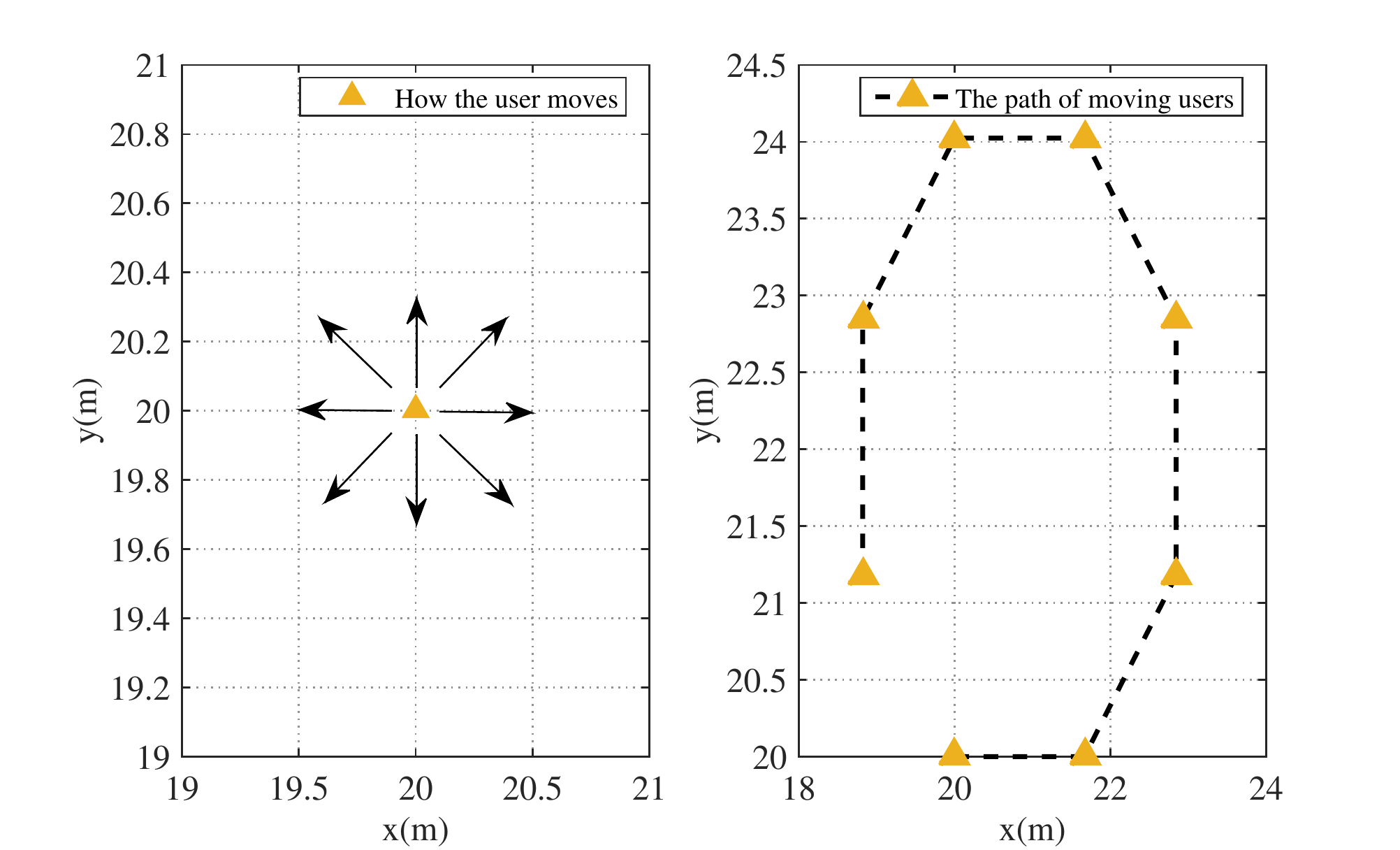}\vspace{-.2cm} 
			\caption{Left hand side: Possible movement directions of moving users. Right hand side: An example of moving user path} \label{fig:2}
		\end{center}
	\end{figure}

	In the previous section, we assumed that the network users are fixed and their locations do not change over the communication time interval $T$. Here, we consider the case of mobile users where users move over the considered coverage area. The user's  movement is assumed to be random obeying the random way point [RWP] \cite{21}. For each user $u_g$ in the multicast group $g$, and in each time slot $n$, we define the $(\mathbf{r}_{g,u_g}[n],\theta[n],V[n])$ where $\mathbf{r}_{g,u_g}[n]$ is the location of the user at the beginning of the time slot $n$, $\theta[n]$ is the movement direction, and $V[n]$ is the user speed. We assume that each user in each time slot moves along the directions from the set $\vartheta=\{0,\pi/4,\pi/2,3\pi/4,\pi,5\pi/4,3\pi/2,7\pi/4\}$ which are selected with uniform distribution. In addition, the speed of users over each time slot is selected from the set $V_\text{user}=[ v1_\text{user},v2_\text{user}]$ with uniform distribution. The movement directions and an example of the user path are shown in Fig. \ref{fig:2}.

	The general model considered is as follows. At the beginning of each time, UAV lies at breaking point $n-1$ with location $\mathbf{q}[n-1]$ and each user lies at its positions determined by $\mathbf{r}_{g,u_g}[n-1]=[\ x_{g,u_g}[n-1],y_{g,u_g}[n-1]]^\text{T}\in \mathbb{R}^{2 \times 1}$. Then we assume that each user changes its location to position $\mathbf{r}_{g,u_g}[n]=[\ x_{g,u_g}[n],y_{g,u_g}[n]]^\text{T}$ based on the random walk point as described in the previous paragraph, and is assumed to be unchanged over time slot $n$. This position can be obtained from the current location of users, their movement direction, and their speeds. Given these new locations of users, during the time slot $n$, UAV moves to the next breaking point, i.e., breaking point $n$ with location $\mathbf{q}[n]$, and consumes transmit power of $p_g[n]$ for information transmission to multicast group $g$. The transmission rate towards the users of this group can be obtained from \eqref{achievable multicast capacity} where the channel gain from UAV to user $u_g$ of multicast group $g$ is given by
	\begin{align}\label{pathlossmodelmoving}
	h_{g,u_g}[n]= \frac{\mu_0}{H^2+{\Vert \mathbf{q}[n] - \mathbf{r}_{g,u_g}[n] \Vert}^2},
	\end{align}
	where by $\mathbf{r}_{g,u_g}[n]$, we explicitly show that the position of users for each time slot could be different.

	Generally, for this kind of consideration, i.e., mobile users, one should use mathematical models with dynamic time adaptation like Markov decision problems (MDP) \cite{17} and machine learning methods \cite{19}. However, in this paper, we consider different scenarios and leave the MDP and learning formulation as future works.

	\subsection{\textcolor{black}{Mobile User Location in ofline mode} \label{Mobile User Location with known user location}
	}
	
	\textcolor{black}{
		In this section, we consider the case where the users are mobile in ofline mode, and the location of users is known at the beginning of communication time. It means that, their locations i.e., $\mathbf{r}_{g,u_g}[n],~ \forall u_g,~g,~n$, over the communication interval, i.e., $T$, it is predictable and known at the beginning of the communication time. This scenario can be considered as the case where one knows the moment movement model of users and adopts an advanced estimation method to estimate the position of users.}
	
	
	The optimization problem, in this case, is the same as the optimization problem \eqref{optimizationproblemmain} except that the user location in each time interval changes. Note that the same solution algorithm as for the optimization problem \eqref{optimizationproblemmain} can be used.

	\subsection{\textcolor{black}{Mobile User Location in online mode}\label{Mobile user location with unknown user location}
	}
	\textcolor{black}{	Online mode is one of the most useful scenarios related to the UAV trajectory topic, that has not been mentioned so far so good. In this scenario, the UAV does not know the location of the users during the communication time at the beginning of the flight.	Knowing the position of users in advance is very hard and advanced estimation techniques cannot achieve the accurate estimation of user positions.\\
		To solve this challenge, we have divided the communication time, into a number of time slots, and according to the position of the users at the beginning of each time slot $n$, the UAV should determine the transmit power for sending signals to users of multicast group $g$, i.e., $p_{g}[n]$, and the location of the next breaking point, i.e., $\mathbf{q}[n]$, where we note that the position of the UAV at the beginning of the time slot $n$ is $\mathbf{q}[n-1]$.}

	
	
	Note that the UAV should finally arrive at breaking point $N$, i.e., $\mathbf{q}[N]$ which is known and fixed. However, as the UAV determines the next location given the current location, the calculated next location can be such far that the UAV could not arrive at $\mathbf{q}[N]$ at the end of communication interval $T$. Therefore, at each time slot, we should limit the set of candidate points for being the next breaking point $\mathbf{q}[n]$ such that the UAV can arrive at the final breaking point during the remaining time for the communication time interval. Defining $\mathnormal{S}_\text{remain}[n] = \frac{\mathnormal{V}_\text{max}T}{N}(N-n),~~n=1,\cdots,N-1$ as the distance which can be traveled by the UAV when it is at breaking point $n$, \textcolor{black}{proposed }the following constraint should be included such that we ensure that UAV will arrive at the final point at the end of the communication interval $T$:
	\begin{align}
	{\Vert \mathbf{q}[n] - \mathbf{q}[N] \Vert}^2 \leqslant {S^2_\text{remain}[n]}.
	\end{align}

	Therefore, for each time slot $n=1,\cdots,  N-1$, the following optimization problem should be solved:
	\begin{subequations}\label{optimizationproblemmainMULUUL}
		\begin{align}
		\max\limits_{\mathbf{p}[n],\mathbf{q}[\mathnormal{n}]}&
		\sum\limits_{g \in \mathcal{G}}C_g[n],~~n=1,\cdots,N-1,
		\\
		\label{optimizationproblemmatext1MULUUL}
		\text{s.t.}~~&{\Vert \mathbf{q}[n] - \mathbf{q}[n-1] \Vert}^2 \leqslant S^2_\text{max},
		\\
		\label{qmobile}
		&{\Vert \mathbf{q}[n] - \mathbf{q}[N] \Vert}^2 \leqslant {S^2_\text{remain}[n]},
		\\
		\label{optimizationproblemmasum1MULUUL}
		&
		C_g[n] \geqslant C_g^{\text{rsv}}[n],
		~~\forall g \in \mathcal{G},
		\\
		\label{optimizationproblemmatwosumMULUUL}
		&
		\sum\limits_{g \in \mathcal{G}}p_{g}[n]\leqslant P_{\text{max}},
		\end{align}
	\end{subequations}
	where we can go through the same steps as for optimization problem \eqref{optimizationproblemmain} to find a solution. 
	
	For the last time slot, i.e., $N$, the next breaking point is $\mathbf{q}[N]$ which is known. This means that for the last time slot, one should only determine the transmit power of the UAV. Therefore, the following power allocation problem should be solved
	\begin{subequations}\label{optmiationproblempowermain12}
		\begin{align}
		\label{maxprob12}\max\limits_{\mathbf{p}}& \sum\limits_{g \in \mathcal{G}}\ln(1+\frac{p_g[N] \hat{h}_g[N]}{I_g[N]+\sigma^2}),
		\\
		\label{maxprobst12}\text{s.t.}~~&
		\ln(1+\frac{p_g[N] \hat{h}_g[N]}{I_g[N]+\sigma^2}) \geqslant C_g^{\text{rsv}}[N],~~\forall g \in \mathcal{G},	
		\\
		\label{maxprobsum12}&
		\sum\limits_{g \in \mathcal{G}}p_{g}[N] \leqslant P_{\text{max}},
		\end{align}
	\end{subequations}
	where we can adopt the same steps as for the power allocation sub-problem of \eqref{optimizationproblemmain}.

	\section{COMPUTATIONAL COMPLEXITY AND
		CONVERGENCE ANALYSIS}\label{COMPUTATIONAL COMPLEXITY AND CONVERGENCE ANALYSIS}
	
	\subsection{Convergence Analysis}\label{Convergence Analysis}

	Note that the ASM-based algorithm iteratively solves the original problem for different subsets of variables fixing other variables. At the main ASM iteration $t$ and given the values of the optimization variables obtained in the previous iteration $t-1$, i.e., $\mathbf{p}^{t-1}$ and $\mathbf{Q}^{t-1}$, the ASM algorithm first determines the trajectory of UAV in iteration $t$, i.e., $\mathbf{Q}^{t}$. This means that the objective value of the main optimization problem increases as the aim is to maximize the objective functions, and hence, we have
	\begin{align}
	OBJ(\mathbf{p}^t,\mathbf{Q}^{t+1})\geqslant OBJ(\mathbf{p}^t,\mathbf{Q}^t).
	\end{align}
	The same argument could be made for the transmit power variables meaning that we have the following results as well:
	\begin{align}
	OBJ(\mathbf{p}^{t+1},\mathbf{Q}^{t+1})\geqslant OBJ(\mathbf{p}^t,\mathbf{Q}^{t+1}).
	\end{align}
	
	We prove the convergence for the main optimization problem \eqref{optimizationproblemmain}. For other problems, the same arguments can be made.
	
	From the above arguments, at each iteration of the main ASM algorithm, the objective values of the main optimization problem increase, and since we have the transmit power constraint, the objective could not grow unbounded, i.e., it is an upper bound. Therefore, starting from an initial feasible point, the solution algorithm converges to a limited value, and hence, the ASM algorithm convergence is guaranteed. 
	
	\subsection{Computational Complexity}\label{Computational Complexity}
	
	\begin{itemize}
		
		\item \textbf{Computational Complexity of Optimization Problem \eqref{optimizationproblemmain}}: The total complexity of the proposed algorithm could be written as $T^\text{ASM}\times (A + B)$ where $T^\text{ASM}$ is the total number of iterations needed for convergence of the ASM procedure, $A$ is the complexity of the SCA-GP-based procedure for solving Problem \eqref{gptransform}, and $B$ is the complexity of the SCA-D.C.-based procedure for solving Problem \eqref{optmiationproblempowermaingp} shown in Algorithm 1. The complexity of the SCA-GP-based procedure for solving Problem \eqref{gptransform} is given by $A=T^\text{SCAG}\frac{\log (\varkappa_A / (t_0 \rho))}{\log(\zeta_0)}$ where $T^\text{SCAG}$ is the total number of iteration for the SCA-GP-based procedure to converge, $\varkappa_A = N+G(N-1)(U_g+2)$ is the total number of constraints of problem \eqref{gptransform}, $t_0$ is the initial point for approximation of the accuracy of the interior point method (IPM), $\rho$ is the stopping criterion of the IPM, and  $\zeta_0$ is for updating the accuracy of the IPM \cite{16}. The complexity of the SCA-D.C.-based procedure for solving Problem \eqref{optmiationproblempowermaingp} is given by $B=T^\text{SCAD}\frac{\log (\varkappa_B / (t_0 \rho))}{\log(\zeta_0)}$ where $\varkappa_B =N (G+1)$.
		
		\item \textbf{Computational Complexity of Mobile User Scheme When the Location of Users Are Known in Advanced}: The computational complexity, in this case, is the same as the computational complexity of the optimization problem \eqref{optimizationproblemmain}.
		
		\item \textbf{Computational Complexity of Mobile User Scheme When the Location of Users Are Not Known in Advanced}:
		The complexity of the proposed algorithm for solving problem \eqref{optimizationproblemmainMULUUL} in each iteration $n=1,\cdots,N-1$ is given by
		$T^\text{ASM}[n]\times (C [n]+ D[n])$ where $C[n]=T^\text{SCAG}[n]\frac{\log (\varkappa_C / (t_0 [n]\rho[n]))}{\log(\zeta_0[n])}$ with $\varkappa_C =2+G(U_g+2)$, and $D[n]=T^\text{SCAD}[n]\frac{\log (\varkappa_D / (t_0[n] \rho[n]))}{\log(\zeta_0[n])}$ with $\varkappa_D =G+1$. The complexity of solving the optimization problem \eqref{optmiationproblempowermain12} is $T^\text{ASM}[N]\times D[N]$. There, the total complexity of this scheme is given by \big($\sum_{n=1}^{N-1} T^\text{ASM}[n]\times (C[n] + D[n])\big)+ \big(T^\text{ASM}[N]\times D[N]\big)$.

	\end{itemize}

	\section{Simulations and Results}\label{sec:simulation}			
	
	\subsection{The case of Fixed Users}
	
	In this section, we present the simulation results for evaluating our proposed scheme. The network parameters are set as $P_{\text{max}}=2$ W, the radius of the coverage area is $R=50$ m, $\sigma^2=-90$ dB, $H=25$ m, $V_{\text{max}}=10$ m/s, $\mu_0=-30$ dB and the stopping criteria values in algorithm 1 are $\varepsilon_1=10^{-2}$, $\varepsilon_2=10^{-3}$. The users are randomly located in the coverage area with uniform distribution.

\begin{figure}[ht]
		\subfigure[]{
			\includegraphics[width=9 cm , height=5.9 cm]{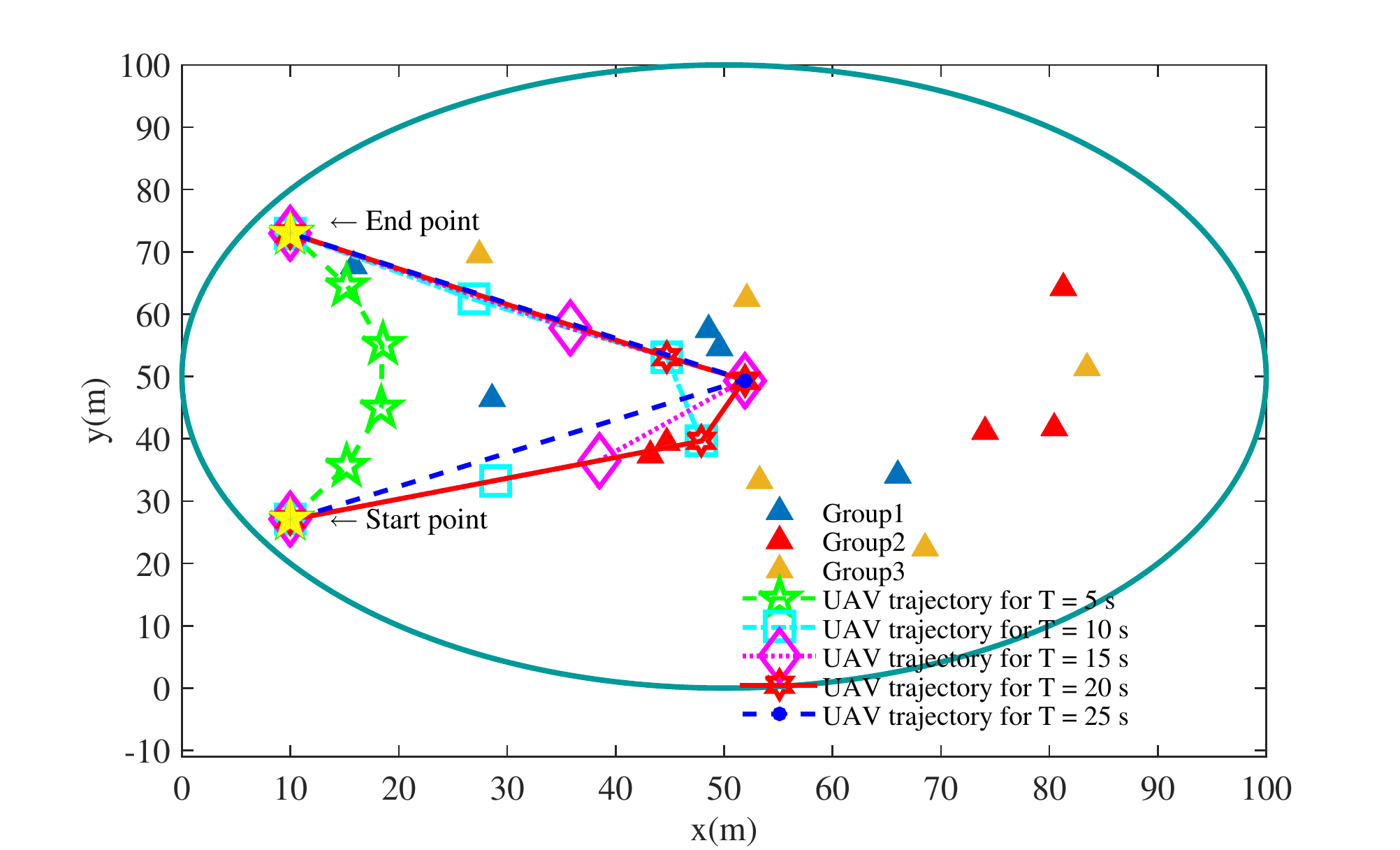}
			\label{fig:3:1}
		}
		\subfigure[]{
			\includegraphics[width=9 cm , height=5.9 cm]{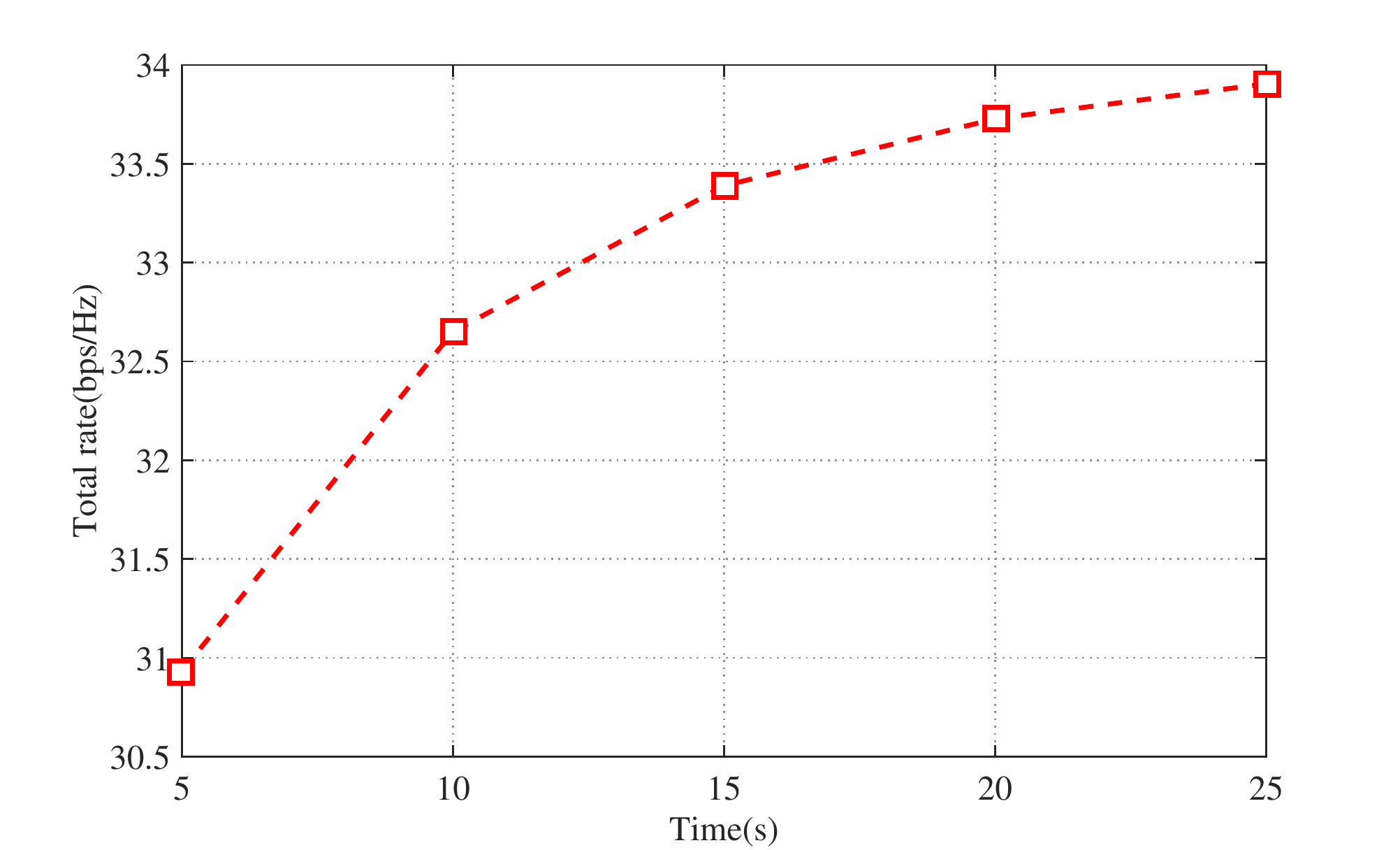}
			\label{fig:3:2}
		}
		\caption{ a) Trajectory of UAV for different values of communications time assuming the same number of breaking points, i.e., $G=3$, $U_g=5$, $N=5$ and $C_g^{\text{rsv}}[n]=0.25$ bps/Hz. b) Total rate vs communication time $T$.}\label{fig:3new}
	\end{figure}

	In the first simulation, we study the effect of communications time on the trajectory of the UAV with fixed numbers of breaking points. We increase the communication time from $\text{T}=5$ s to $\text{T}=25$ s, solve the optimization problem \eqref{optimizationproblemmain}, and plot the results in Fig. \ref{fig:3new}. When the communications time is small, the distance between two consecutive breaking points is also small. Constraint \eqref{optimizationproblemmatext1} enforces the trajectory to be short. However, as the communications time increases, the distance between two consecutive breaking points increases; hence, the UAV can fly to a more distant point in which the UAV could send the information at higher rates. For some values of the communications time, the UAV could reach a point in which the transmission rate is maximum over all the points in the trajectory. Above this time, the UAV spends more time in the maximizing point, i.e., there is more than one breaking point that coincides with this maximizing point. Therefore, as shown in Fig. \ref{fig:3:2}, the total data rate increases when the communication time $T$ increases.
	
\begin{figure}[ht]
		\subfigure[]{
			\includegraphics[width=9 cm , height=5.9 cm]{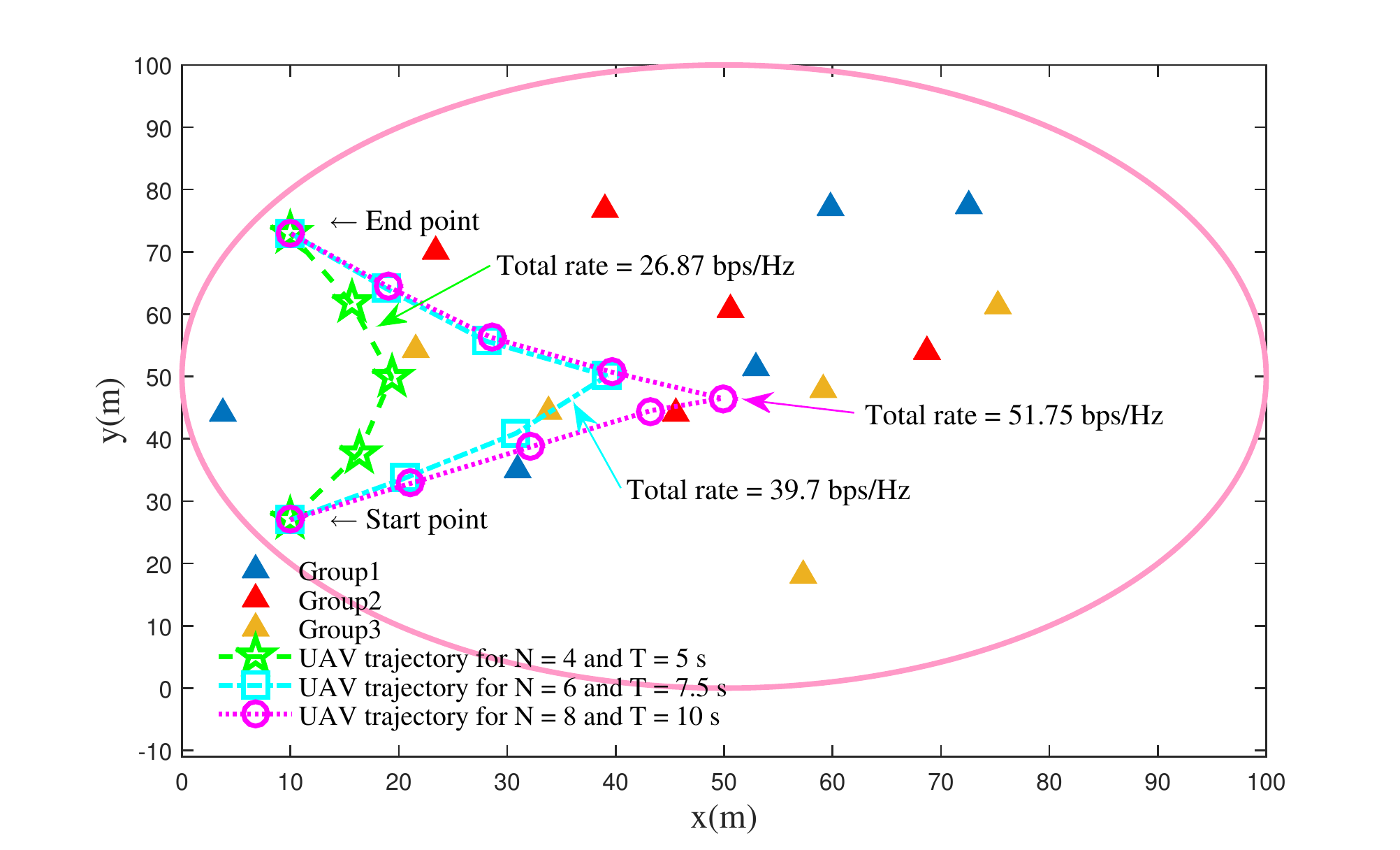}
			\label{fig:5:1}
		}
		\subfigure[]{
			\includegraphics[width=9 cm , height=5.9 cm]{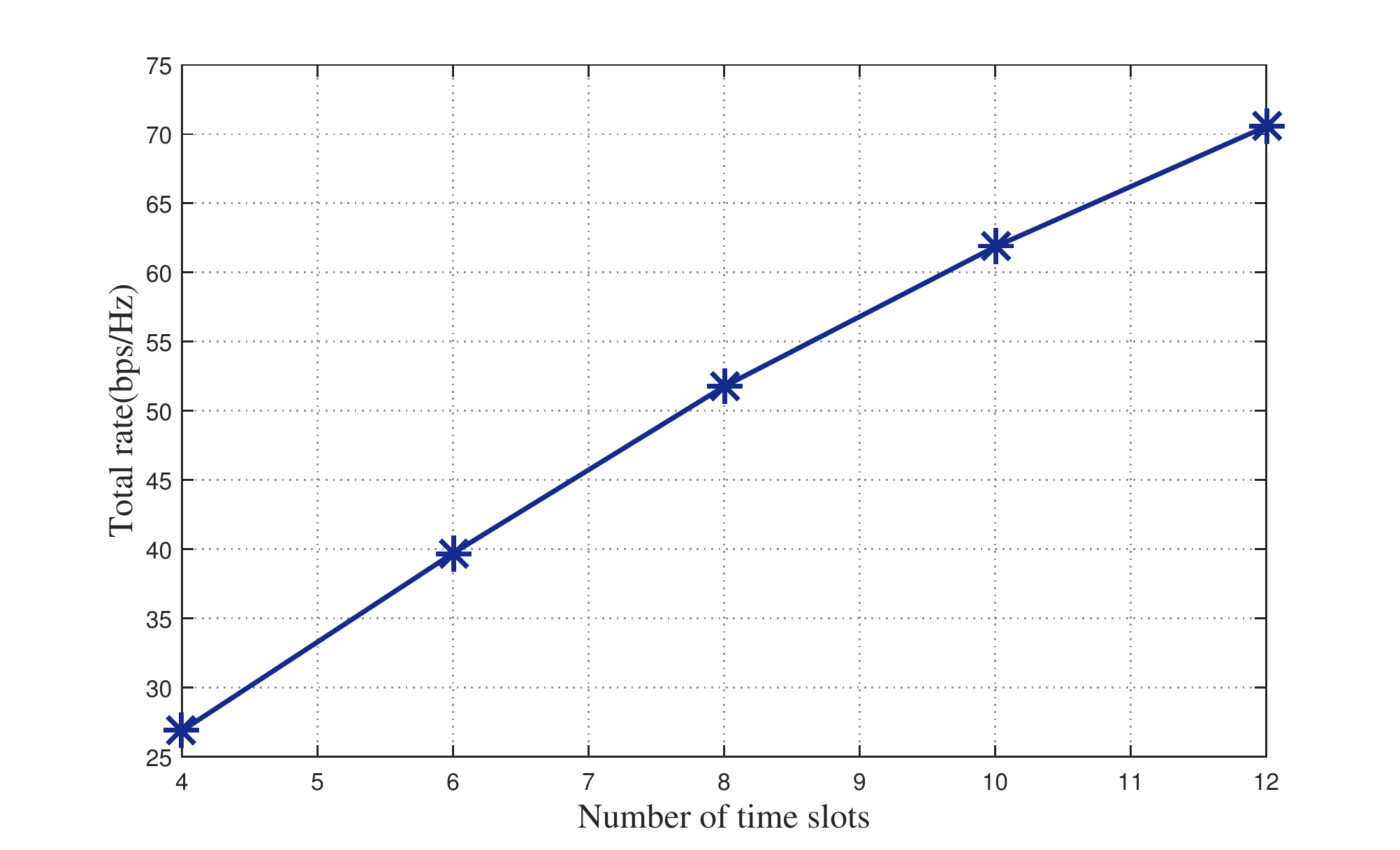}
			\label{fig:5:2}
		}
		\caption{ a) Trajectory of UAV for different values of communications time assuming the same value for the speed constraint, i.e., $G=3$, $U_g=5$,  $C_g^{\text{rsv}}[n]=0.25$ bps/Hz, $S_\text{max}=12.5$ m. 
			\textcolor{black}{ b) Total rate vs number of time slots $N$.}}\label{fig:5new}
	\end{figure}

	We also study the effect of the number of breaking points when the distance between two consecutive breaking points is fixed. This means that we increase the communications time, i.e., $T$, by  $T=\frac{\mathnormal{S}_\text{max}N}{\mathnormal{V}_\text{max}}$ where we assume that the maximum distance between two consecutive breaking points is known and fixed to $\mathnormal{S}_\text{max}$. The result is plotted in Fig. \ref{fig:5:1}. As could be seen, when the number of breaking points is small, the trajectory is short and the UAV does not reach the maximizing point. However, as the number of breaking points increases (at the same time the communications time increases), the UAV could reach the maximizing point and stay moreover this point to a achieve higher data rate.
	\\ 
	\textcolor{black}{
		As shown in Fig. \ref{fig:5:1}, by increasing the number of breaking points, it increases the duration of the UAV's flight, and the communications time, and thus the UAV serves a larger number of users and also achieves higher total rates.
		Now, in the Fig. \ref{fig:5:2}, this issue has been simulated more clearly, and as you can see, by increasing the number of time slots, the UAV serves more users, and therefore the available total rate with fixed $\mathnormal{S}_\text{max}$, has also increased.}
	
\begin{figure}[h]
			\subfigure[$U_g=3, ~\forall g\in\mathcal{G}$]{
				\includegraphics[width=9 cm , height=5.9 cm]{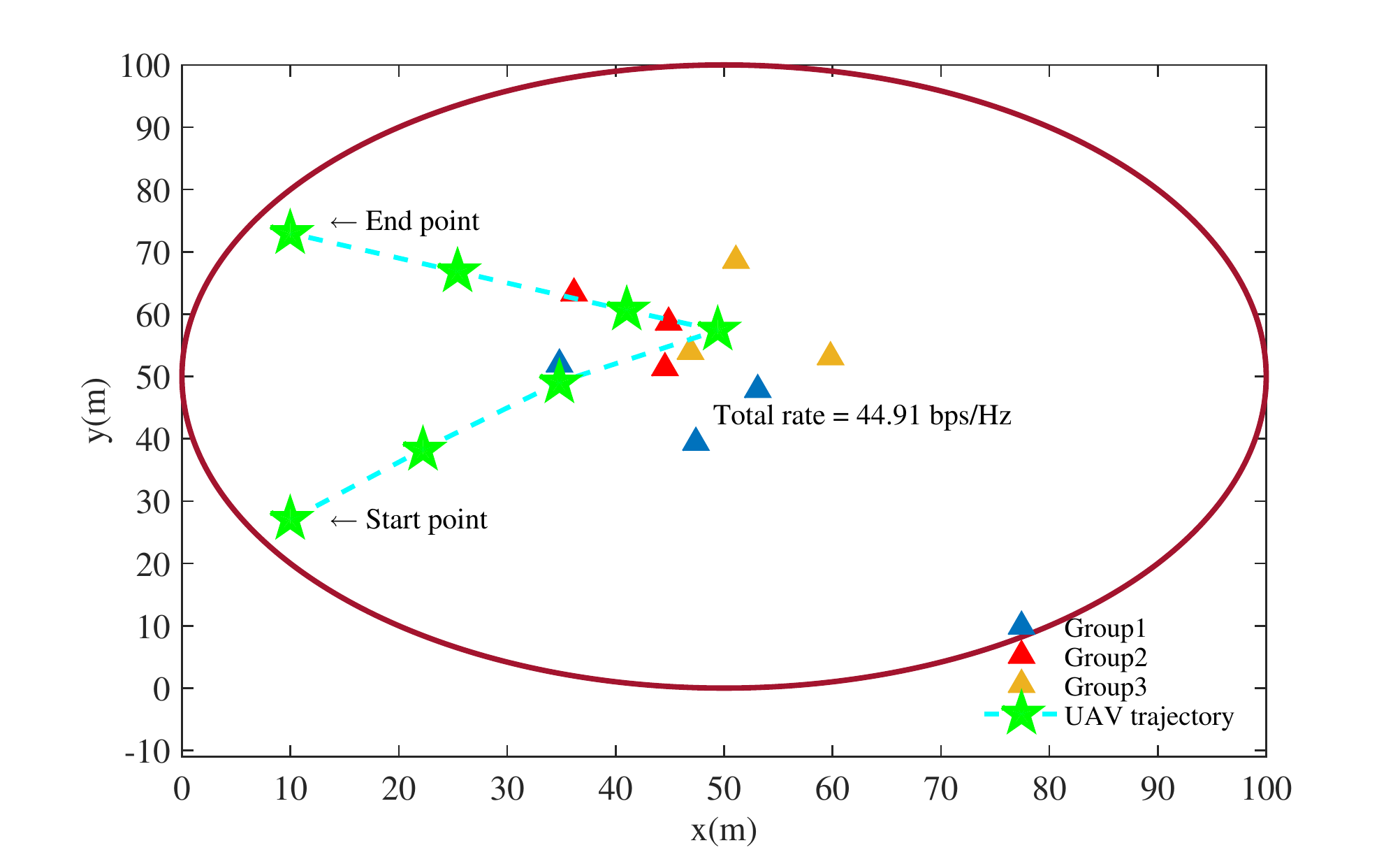}
				\label{fig:y equals}
			}
			\subfigure[ $U_g=5, ~\forall g\in\mathcal{G}$]{
				\includegraphics[width=9 cm , height=5.9 cm]{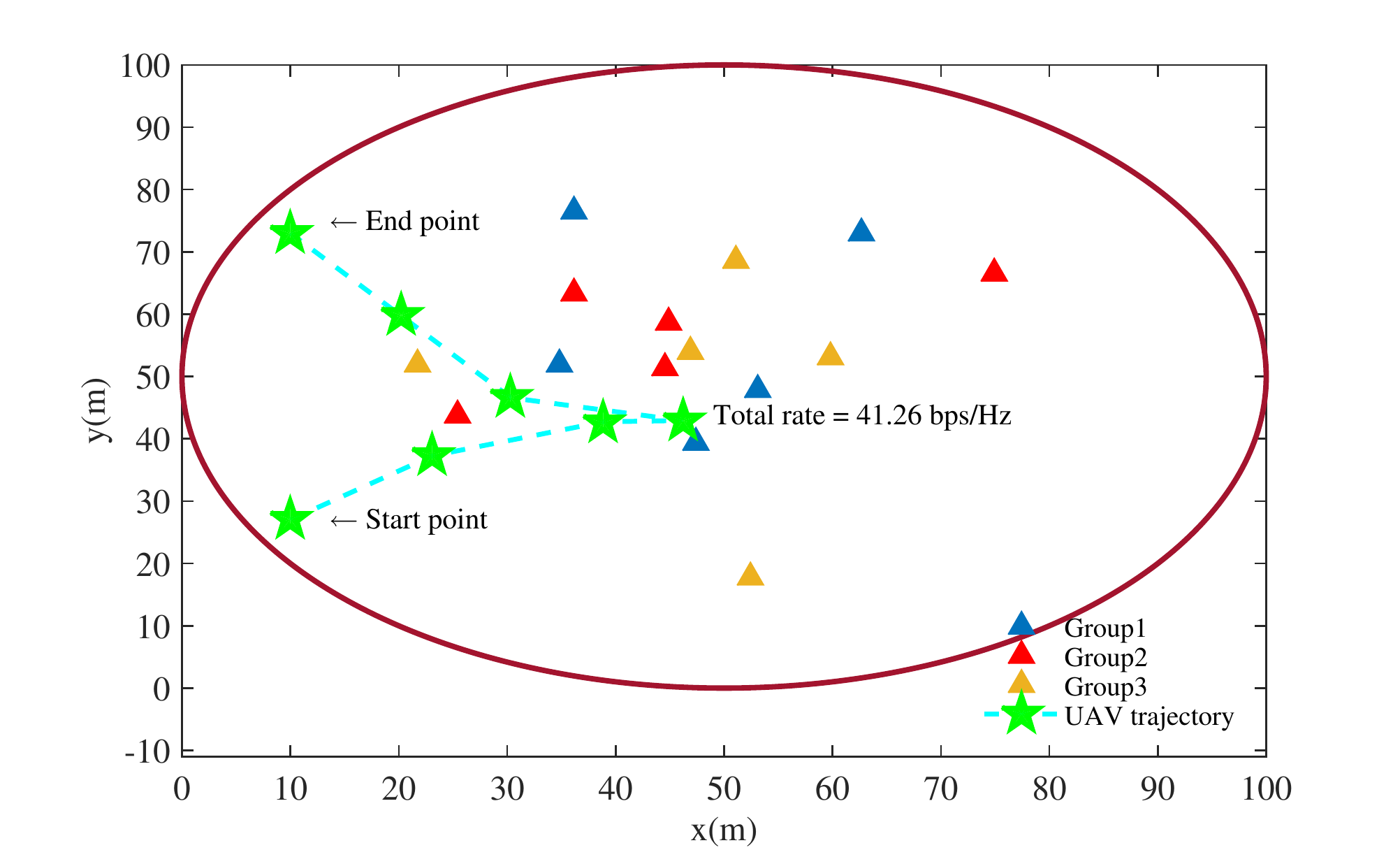}
				\label{fig:three sin x}
			}
			\subfigure[$U_g=7, ~\forall g\in\mathcal{G}$]{
				\includegraphics[width=9 cm , height=5.9 cm]{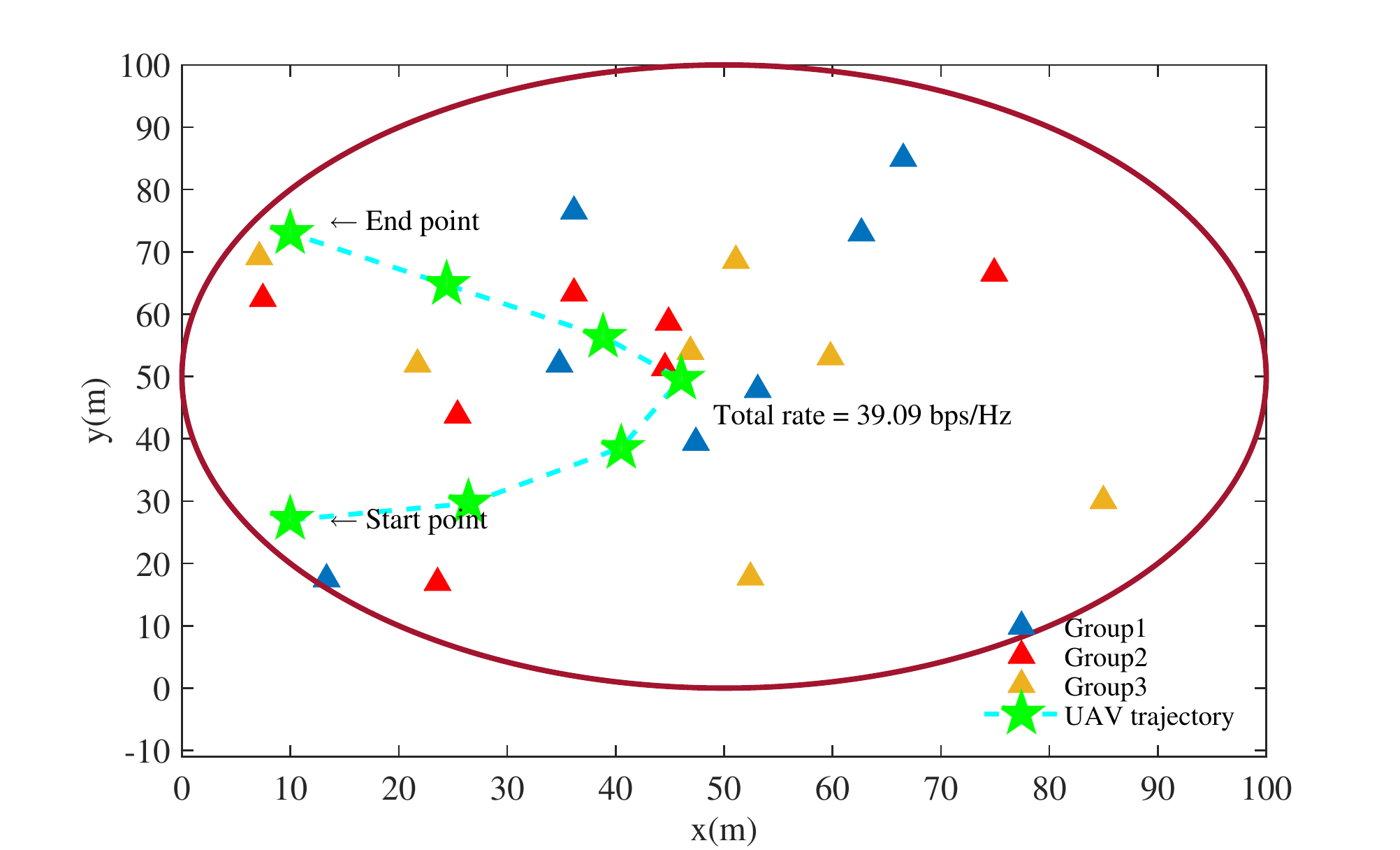}
				\label{fig:five over x}
			}
			\caption{Trajectory of UAV for different number of users per group for $G=3$, $T=10$s, $N=6$, $C_g^{\text{rsv}}[n]=0.25$ bps/Hz. 
			}\label{fig:three graphs}
		\end{figure}

	Next, we study the effect of the number of users in the network. We assume that there exist $G=3$ multicast groups in the network each with the same number of users. We change the number of users in multicast groups and solve the proposed optimization problem to find the transmit powers and the trajectory of the UAV. The result is plotted in Fig. \ref{fig:three graphs}. As could be seen, increasing the number of users in multicast groups decreases the total transmission rate of the network. This is because, in multicast transmission, the data rate of each group is limited by the worst user. Adding new users to the multicast groups would introduce users with worse channel gain compared to the existing users, and hence, the data rate decreases.

\begin{figure}[h]
			\begin{center}
				\includegraphics[width=9 cm , height=5.9 cm]{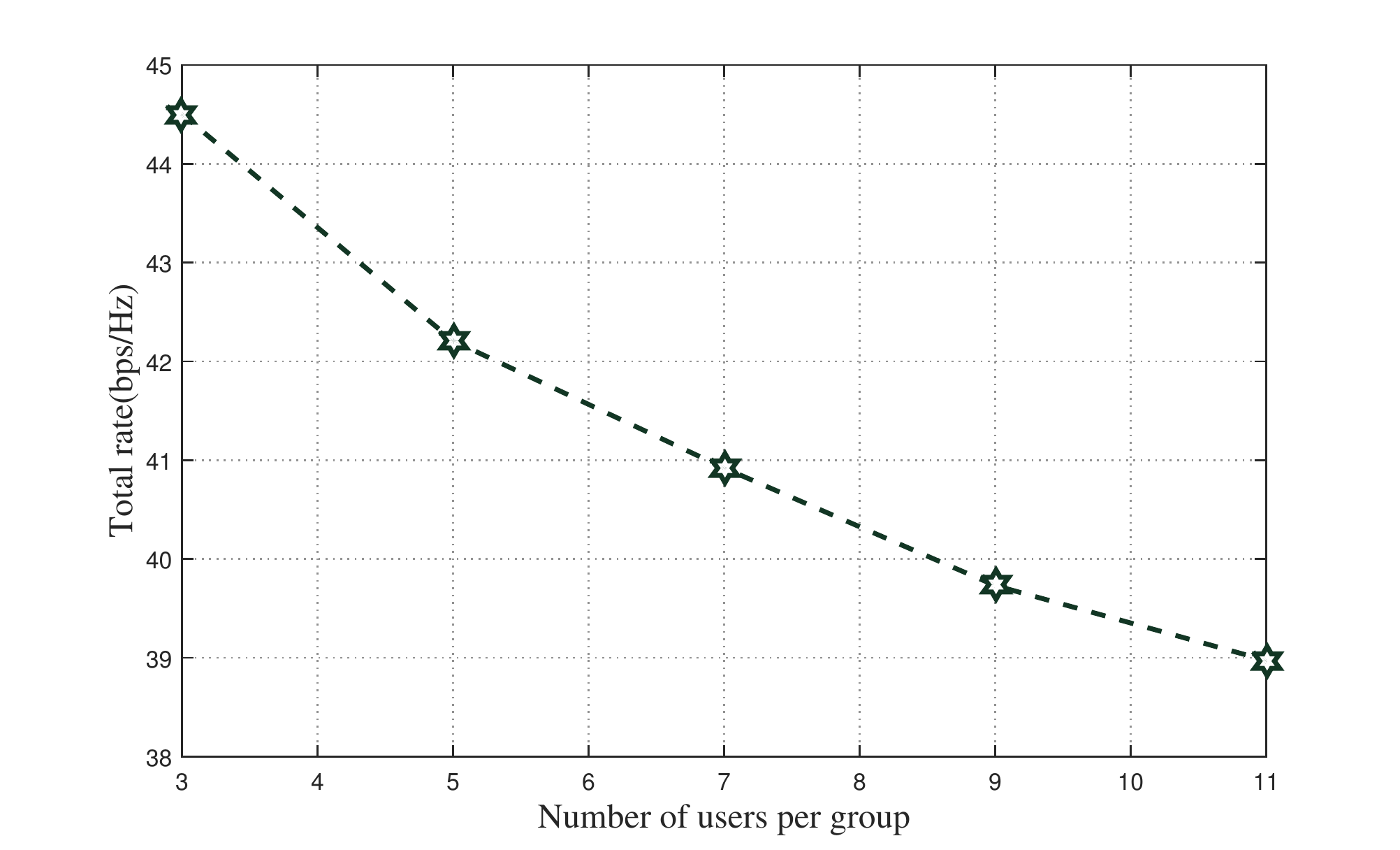}\vspace{-.2cm} 
				\caption{\textcolor{black}{Total rate vs number of users per groups $U_g$ for $G=3$, $T=10$s, $N=6$, $C_g^{\text{rsv}}[n]=0.25$ bps/Hz.} 
				} \label{fig:71}
			\end{center}
		\end{figure}

		\textcolor{black}{
			To show Fig. \ref{fig:three graphs} with another approach, we have shown the total rate vs increasing number of users per group in Fig. \ref{fig:71}. In each run by preserving previous users, we have added new number of users. If the added users based on multicast scenario have better situation than existing users i.e., in terms of channel gain, there is no difference in the total rate but if the added users don't have a good situation the total rate is decreased.
		}

		\begin{figure}[h]
			\subfigure[ $G=3$ and $U_g=4, ~\forall g\in\mathcal{G}$]{
				\includegraphics[width=9 cm , height=5.9 cm]{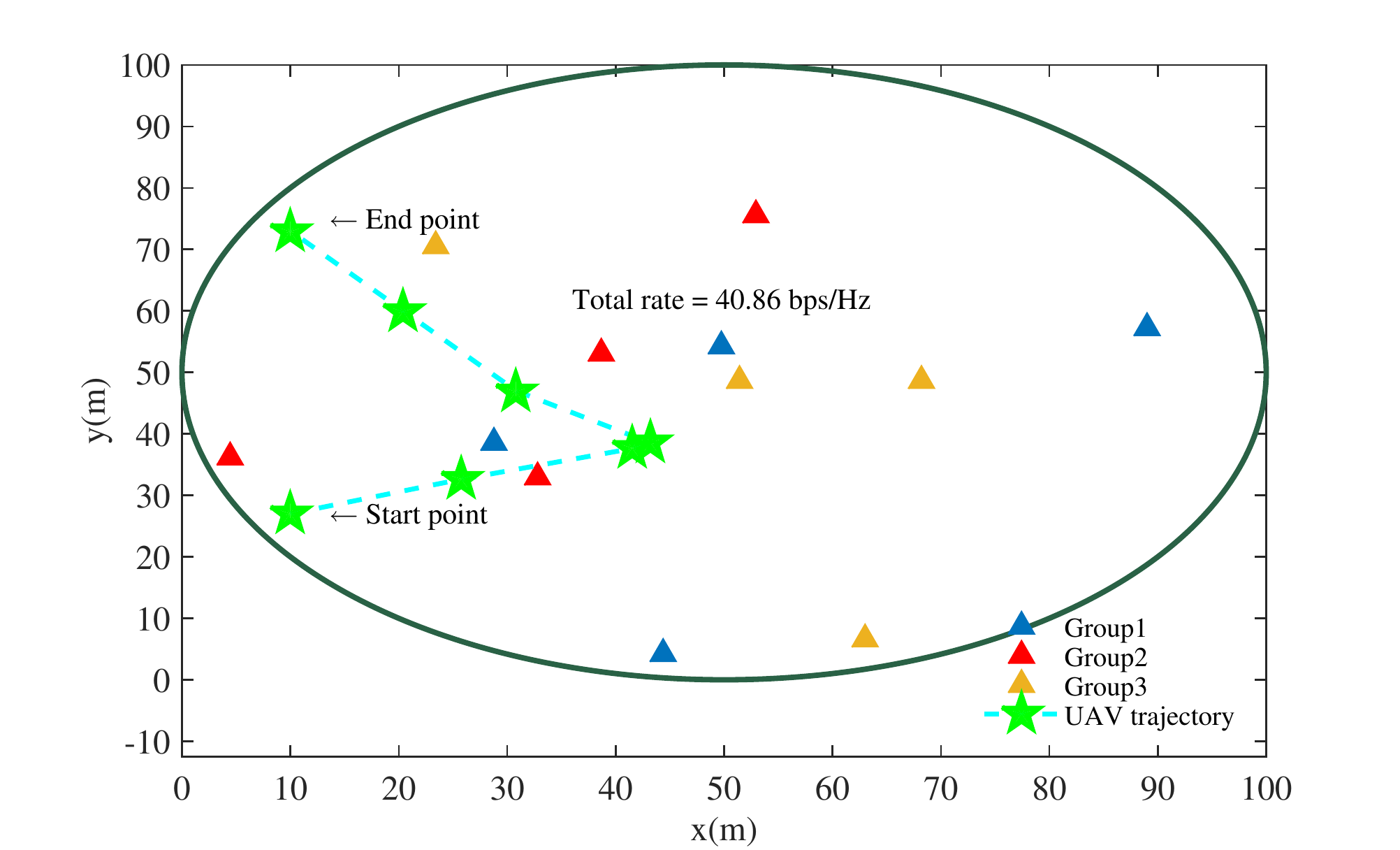}
				\label{fig:y equals x}
			}
			\subfigure[$G=6$ and $U_g=2, ~\forall g\in\mathcal{G}$]{
				\includegraphics[width=9 cm , height=5.9 cm]{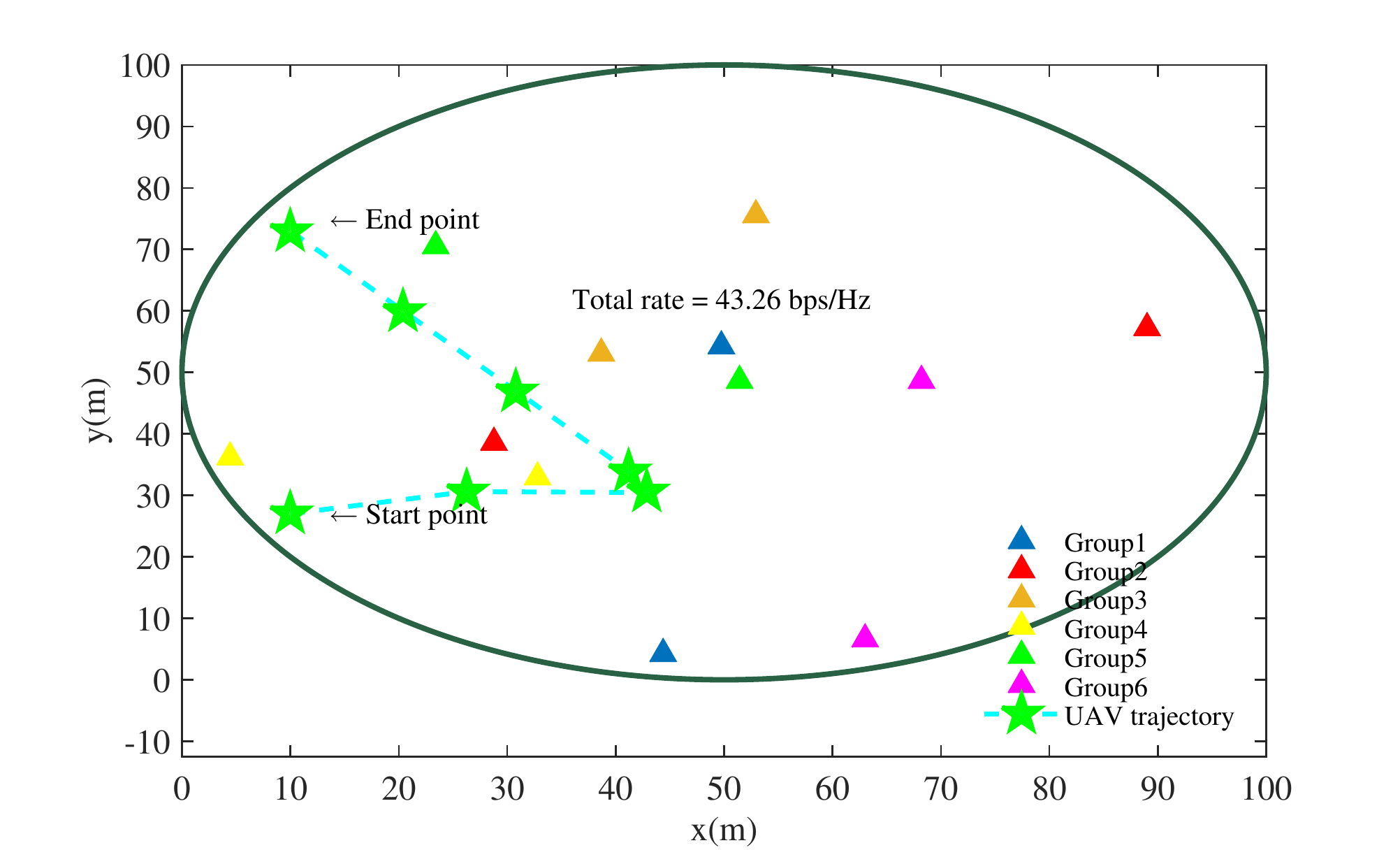}
				\label{fig:unicast1}
			}
			\subfigure[$G=12$ and $U_g=1, ~\forall g\in\mathcal{G}$]{
				\includegraphics[width=9 cm , height=5.9 cm]{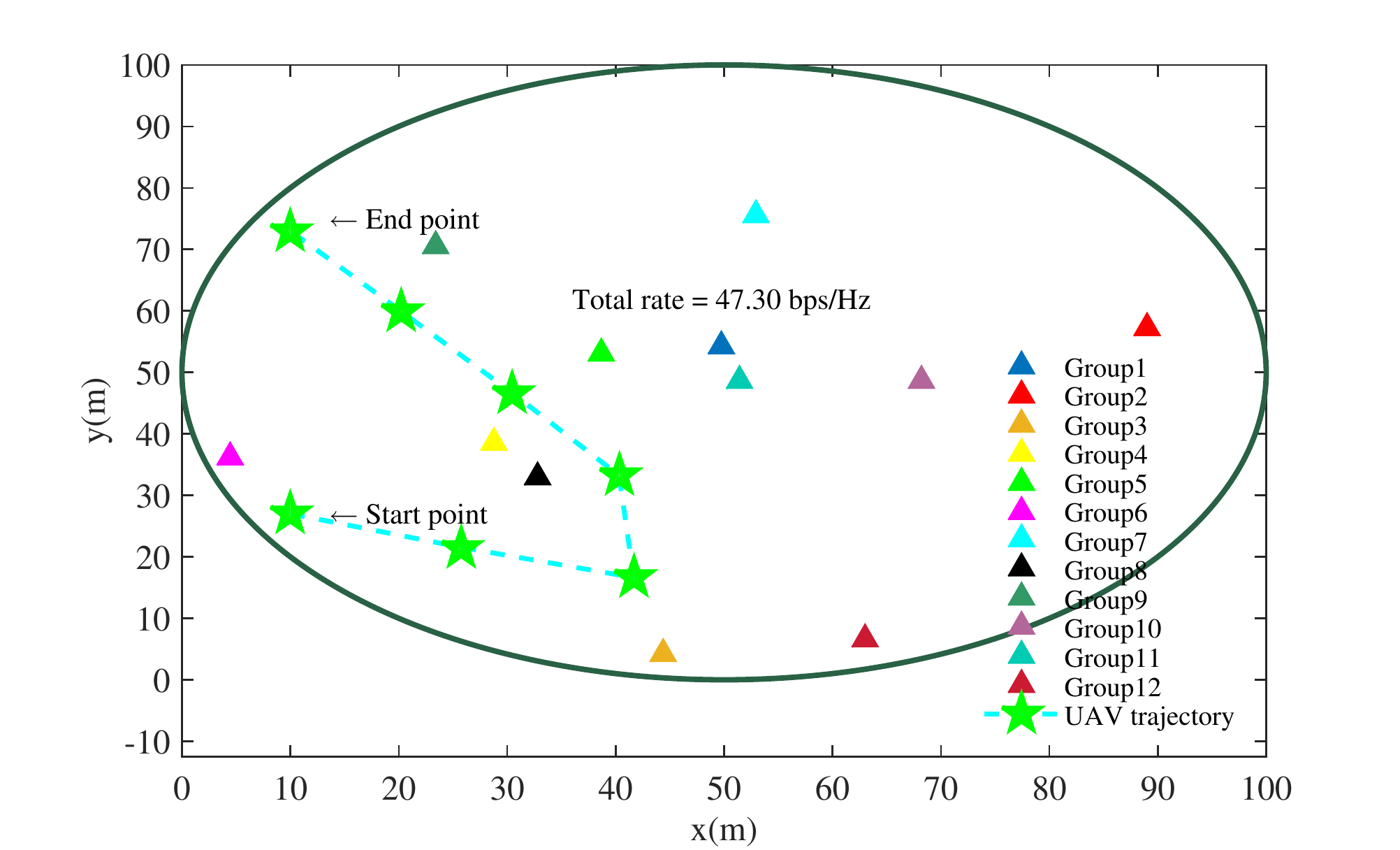}
				\label{fig:unicast2}
			}
			\caption{Trajectory of UAV for different number of groups assuming the same total number of users, i.e., $U=12$, and hence, different number of users per group, and $T=10$s, $N=6$, $C_g^{\text{rsv}}[n]=0.08$ bps/Hz.
			}\label{fig:unicast3}
		\end{figure}

		For the fixed user scenario, we study the effect of the number of multicast groups. \textcolor{black}{The main idea of analyzing this part is to investigate the proposed system between multicast and unicast communication models. The decreasing number of users and an increasing number of groups with the aim of keeping the number of users fixed. We convert the proposed system with three groups that in each one exist four users to twelve groups that exist one user in each group. Hence, we can transform multicast to unicast scenario 
			.} 
		We assumes the fixed number of network users, i.e., 12 users, at first divide them into three groups each with 4 users. Next, we divide the total users into 6 groups each with 2 users. Finally, we have 12 groups each with one user which could be considered, as the unicast scenario. We set the minimum rate requirement to $C_g^{\text{rsv}}=0.08$ bps/Hz. The results are plotted in Fig. \ref{fig:unicast3}. As could be seen, by increasing the number of groups, and hence decreasing the number of users in each group, the total data rate increases. This is due to the fact that in multicast transmission, the worst user in each group is the bottleneck; hence by splitting a group into two or more groups, the users which are bottleneck in each group would have better channel conditions, and hence, the total data rate of the network will increase. 
		
		\subsection{The case of Mobile Users}
		
		Now, we provide simulation results for our proposed scheme for the case of mobile users. The velocity of users is chosen from interval $V_\text{user}=[ 0 , 3 ]$ with uniform distribution. 
		In the first simulation which is for the case where the location of users is known a priori, we run our proposed solution algorithm for the corresponding optimization problem, i.e., the optimization problem \eqref{optimizationproblemmain} in which the location of users are considered to change over each time slot, and plot the results in Fig. \ref{mobileuser}. Note that in this case, solving the optimization problem results in finding the location of the UAV in each time slot, and hence we have the whole trajectory of the UAV before communication starts. This trajectory is shown in Fig. \ref{mobileuser7g}. However, Figs. \ref{mobileuser7a} to \ref{mobileuser7f} show the time lapse of the communication scenario. In Fig. \ref{mobileuser7a}, the UAV is at its start point and users are in their new locations which are fixed over time slot $n=1$. In Fig. \ref{mobileuser7b}, the UAV flies in the next breaking point in time slot 2 and the users have moved to their new locations. Note that, as the users are mobile, we had to plot these figures where in each figure, the locations of users are different from other figures.

\begin{figure}[h]
			\begin{center}
				\includegraphics[width=9 cm , height=5.9 cm]{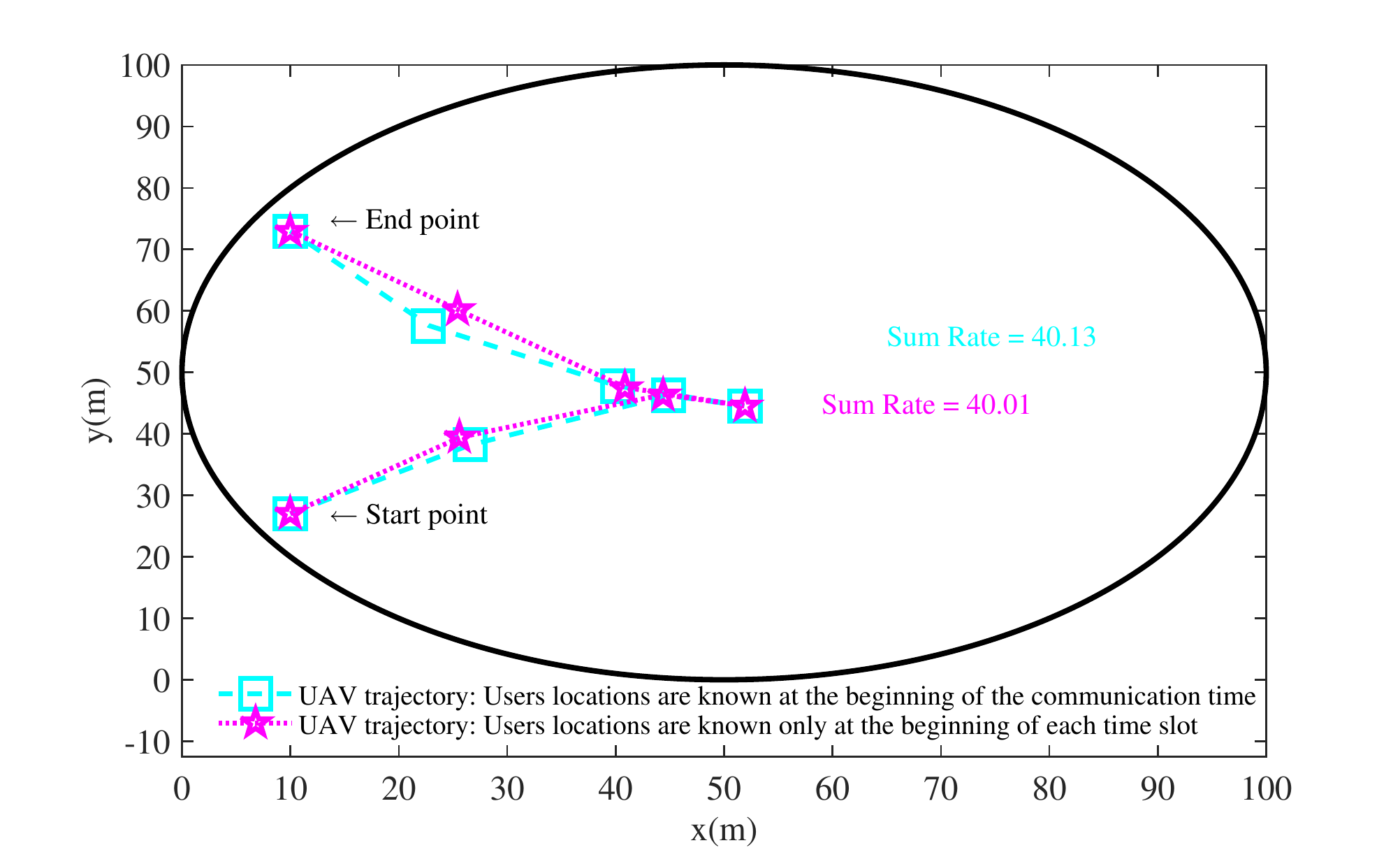}\vspace{-.2cm} 
				\caption{The comparison between the trajectories of the UAV when the locations of users are known at the beginning of the communication time interval and when this information are known at the beginning of each time slot. System parameters are $G=3$, $T=12$s, $N=6$, $C_g^{\text{rsv}}[n]=0.25$ bps/Hz, $U_g=5, ~\forall g\in\mathcal{G}$.} \label{Fig8}
			\end{center}
		\end{figure}

		We next provide results for the case where the locations of users are not known at the beginning of the communication interval as shown in Fig. \ref{Fig8}. For the sake of comparison, we also provide the results for the case when the locations of users are known in advance. As can be seen, the trajectory obtained by the proposed method in the case of no prior user location information is the same as that in the case where the user locations are known in advance. In addition, the total achieved data rate of both cases is the same, and we note that the UAV has arrived at the final location for both cases.

\begin{figure}[h]
			\begin{center}
				\includegraphics[width=9 cm , height=5.9 cm]{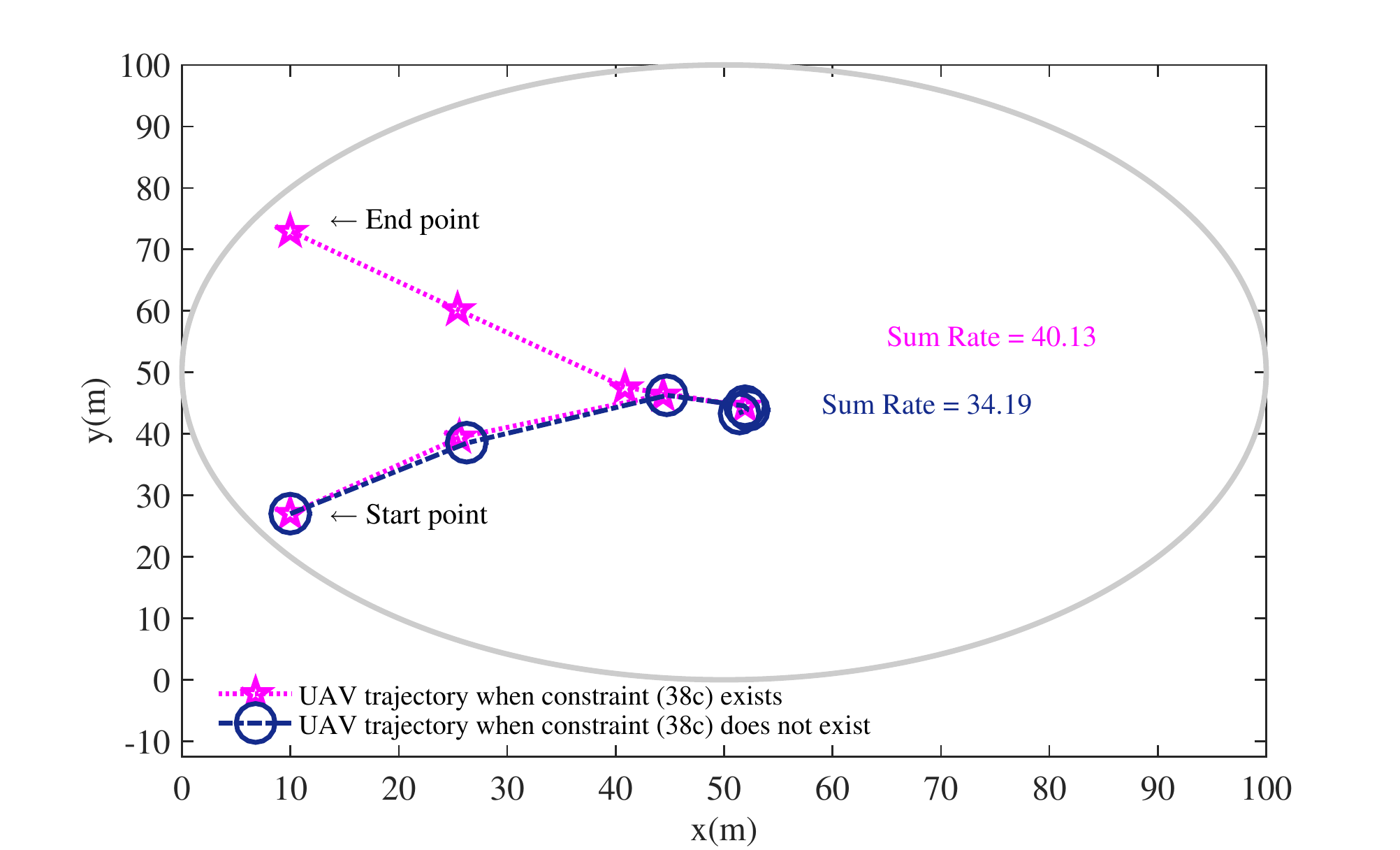}\vspace{-.2cm} 
				\caption{The comparison of the trajectories of the UAV when the locations of users are known only at the beginning of each time slot for cases when constraint  \eqref{qmobile} is absent in the optimization problem \eqref{optimizationproblemmainMULUUL} and when this constrain is included in the optimization problem \eqref{optimizationproblemmainMULUUL}. System parameters are $G=3$, $T=12$s, $N=6$, $C_g^{\text{rsv}}[n]=0.25$ bps/Hz, $U_g=5, ~\forall g\in\mathcal{G}$.} \label{Fig9}
			\end{center}
		\end{figure}

		In Fig. \ref{Fig9}, investigate the necessity of constraint \eqref{qmobile}. As we stated before, this constraint is needed to derive the UAV to its final location. In the absence of \eqref{qmobile}, the UAV may move to a location that is optimal, i.e., maximum transmission rate, and stay over that point for all the remaining time. This is because, for each time slot, a different optimization problem is solved whose solution is the next location of the UAV which maximizes the transmission rate. As seen in Fig. \ref{Fig9}, when constraint \eqref{qmobile} is not included, the UAV does not move towards its final location. However, as including this constraint makes the UAV choose its next location with the view of reaching its final location, the UAV will arrive at the final location at the end of communication time, i.e., time slot $N$.
		
\begin{figure}[h]
			\subfigure[Example of user distribution in time slot 1]{
				\includegraphics[width=9 cm , height=5.9 cm]{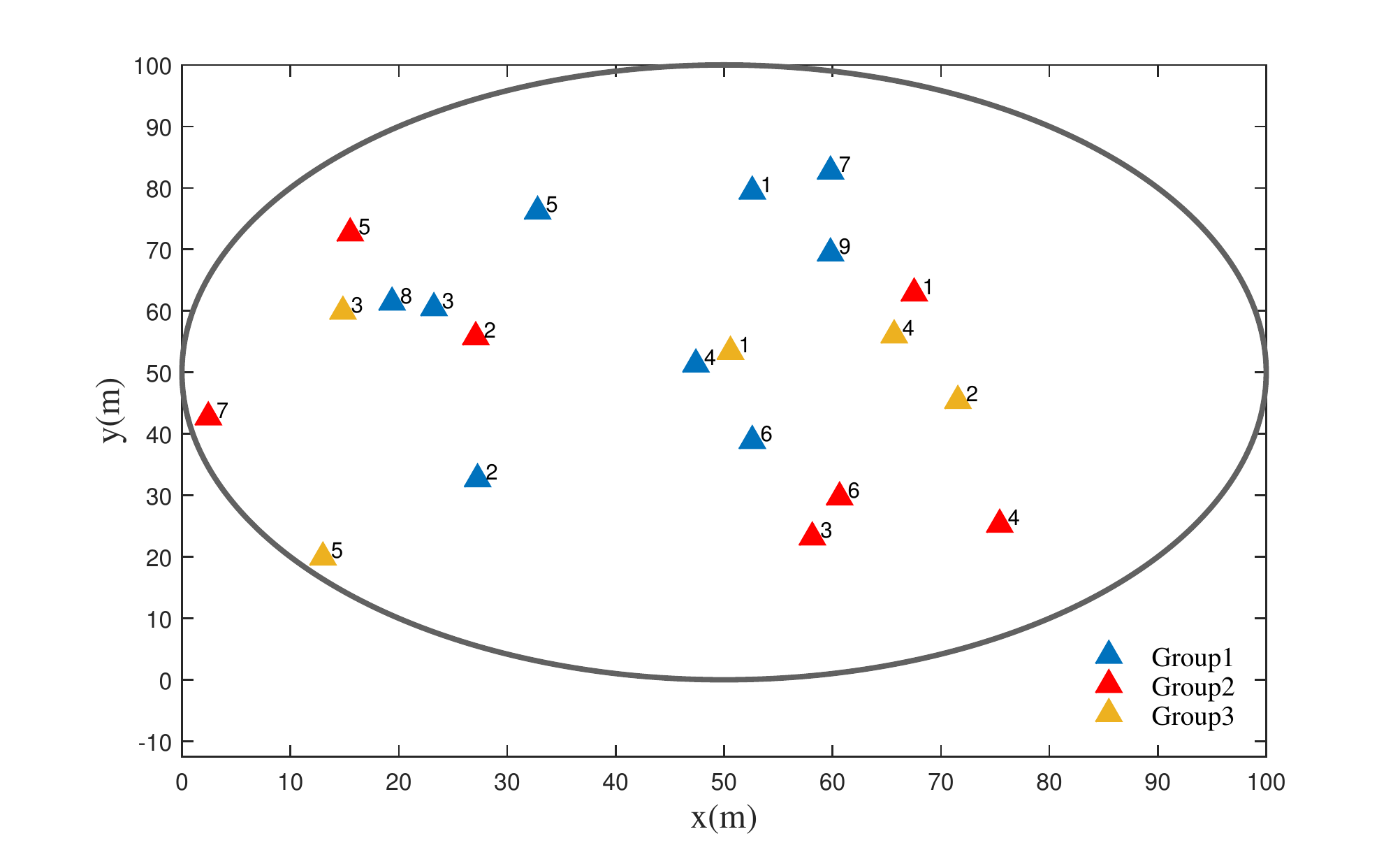}
				\label{fig:13a}
			}
			\subfigure[Example of user distribution in time slot 2]{
				\includegraphics[width=9 cm , height=5.9 cm]{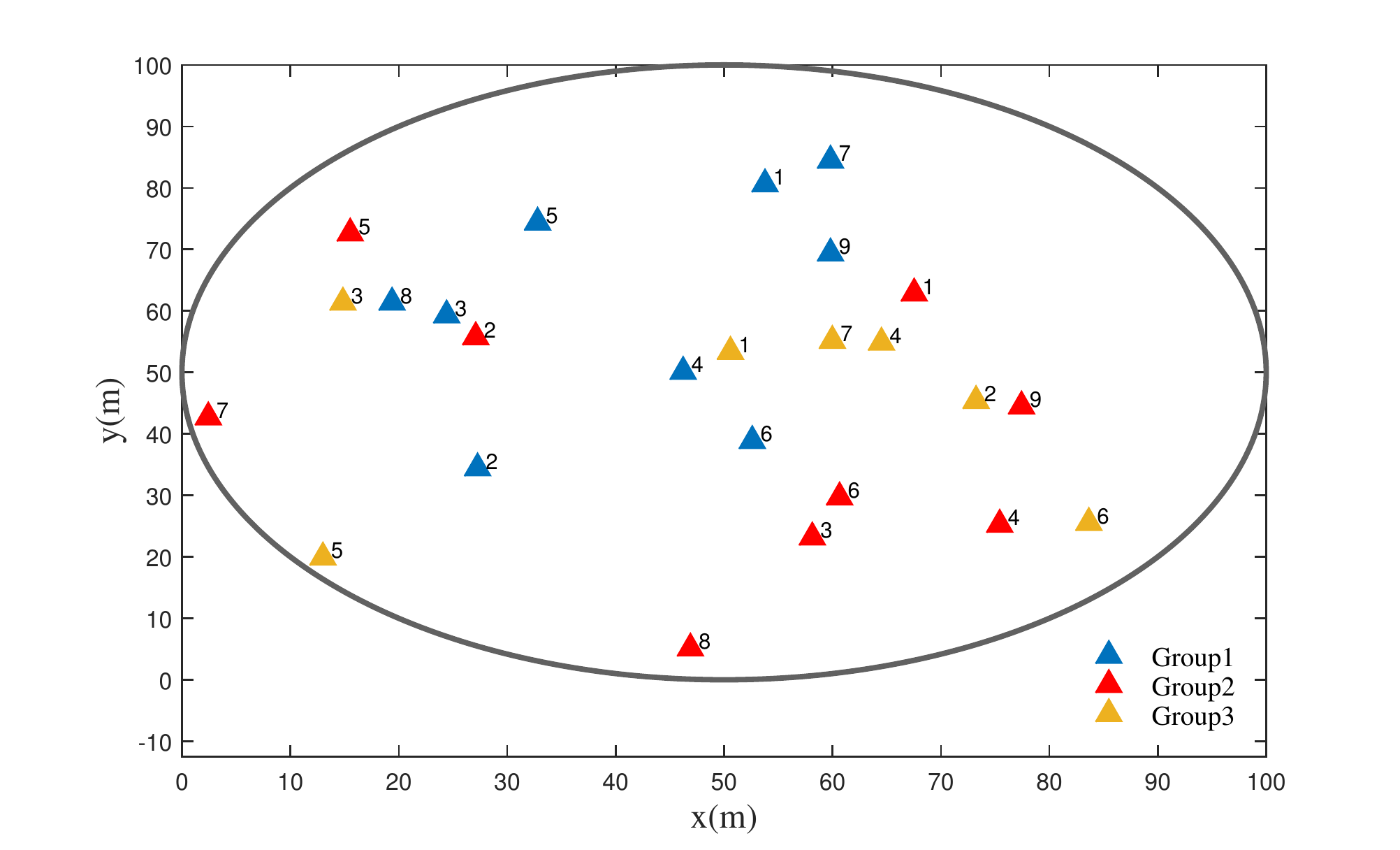}
				\label{fig:13b}
			}
			\subfigure[The variation of rate versus power and $\lambda$ ]{
				\includegraphics[width=9 cm , height=5.9 cm]{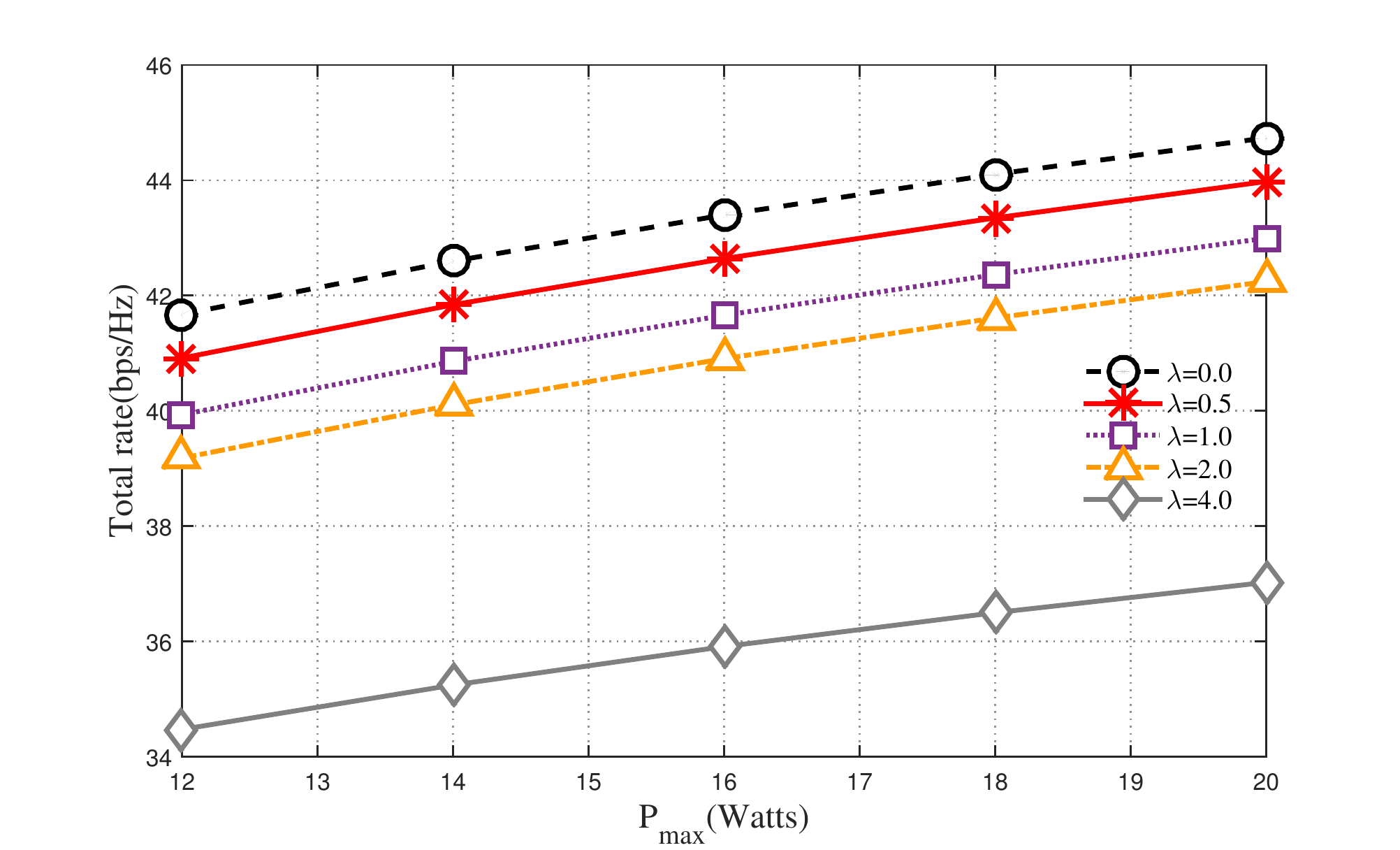}
				\label{fig:13c}
			}
			\caption{The effect of changing the number of mobile users on online scenario.  (a), (b) Show user’s placement in two different time slots, (c) The variation of total rate vs power for different value of $\lambda$, i.e., $G=3$, $T=12s$,  $C_g^{\text{rsv}}[n]=0.25$ bps/Hz, $N=6$.}
			\label{fig:13new}
			\end{figure}

\begin{figure*}[h]
	\normalsize
	\centering  
	\subfigure[Users positions and UAV trajectory at the beggining time slot 1
	]
	{\includegraphics[width=0.495\linewidth]{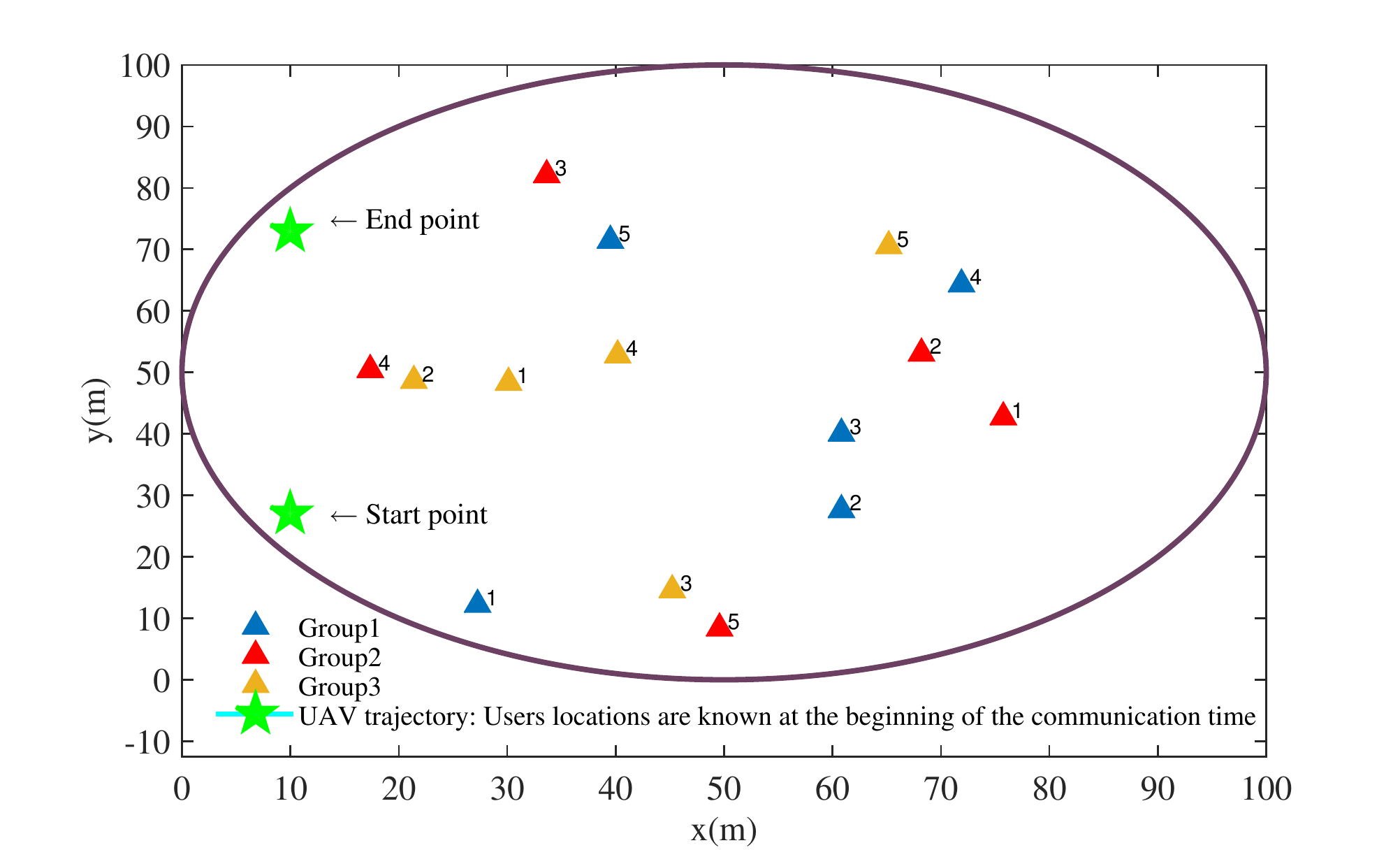}\label{mobileuser7a}}
	\subfigure[Users positions and UAV trajectory at the beggining time slot 2
	]
	{\includegraphics[width=0.495\linewidth]{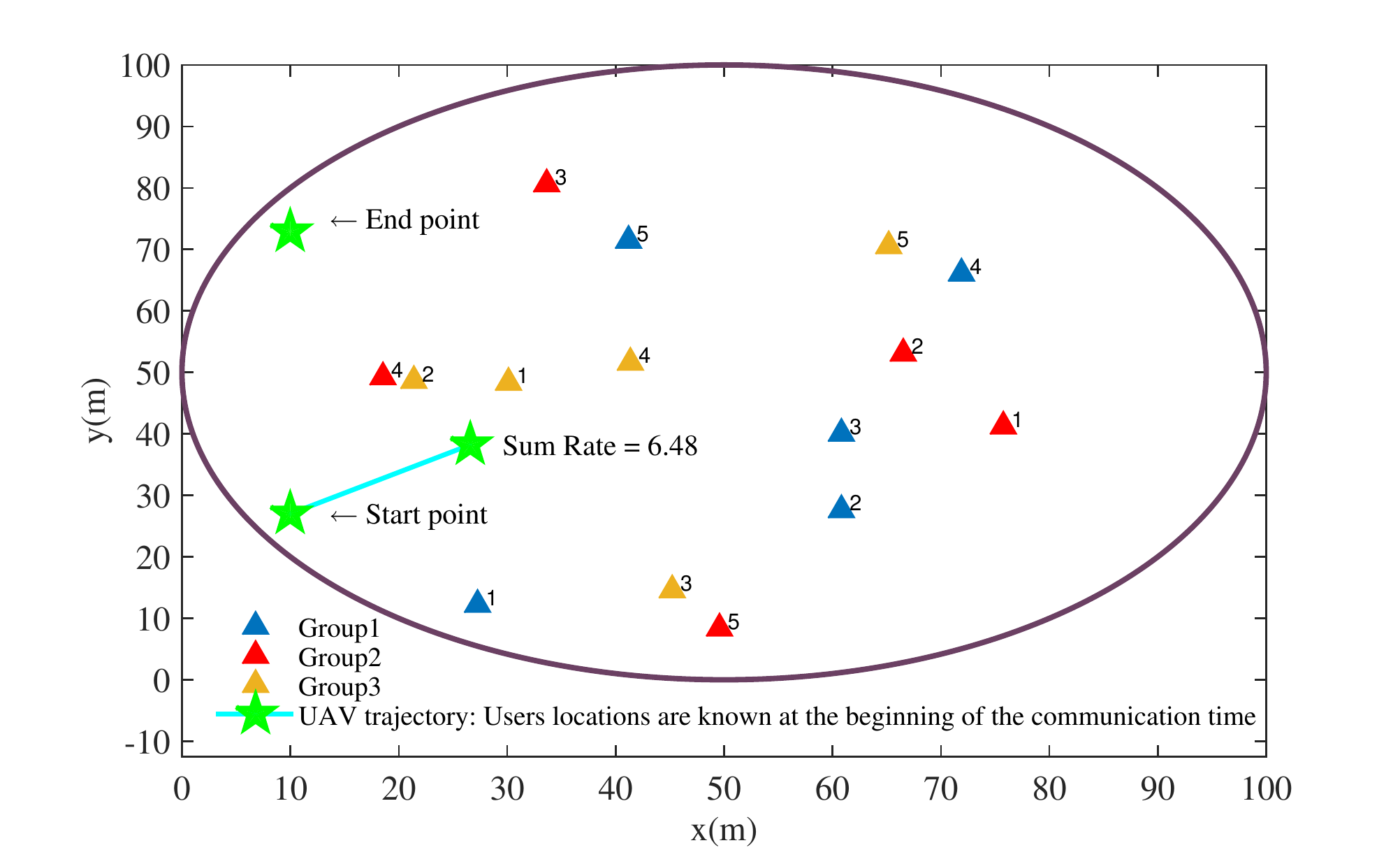}\label{mobileuser7b}}
	\subfigure[Users positions and UAV trajectory at the beggining time slot 3
	]
	{\includegraphics[width=0.495\linewidth]{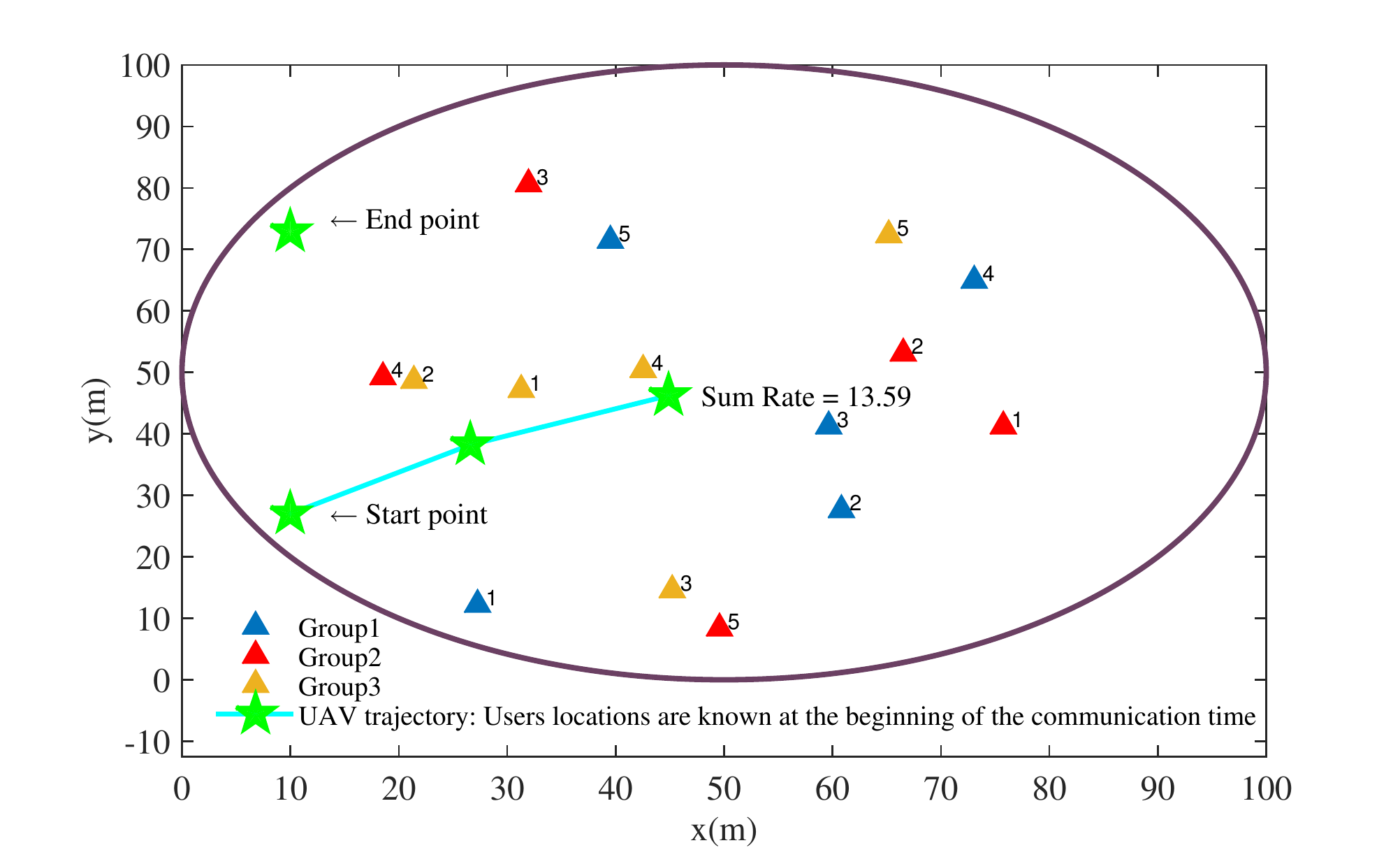}\label{mobileuser7c}}
	\subfigure[Users positions and UAV trajectory at the beggining time slot 4
	]
	{\includegraphics[width=0.495\linewidth]{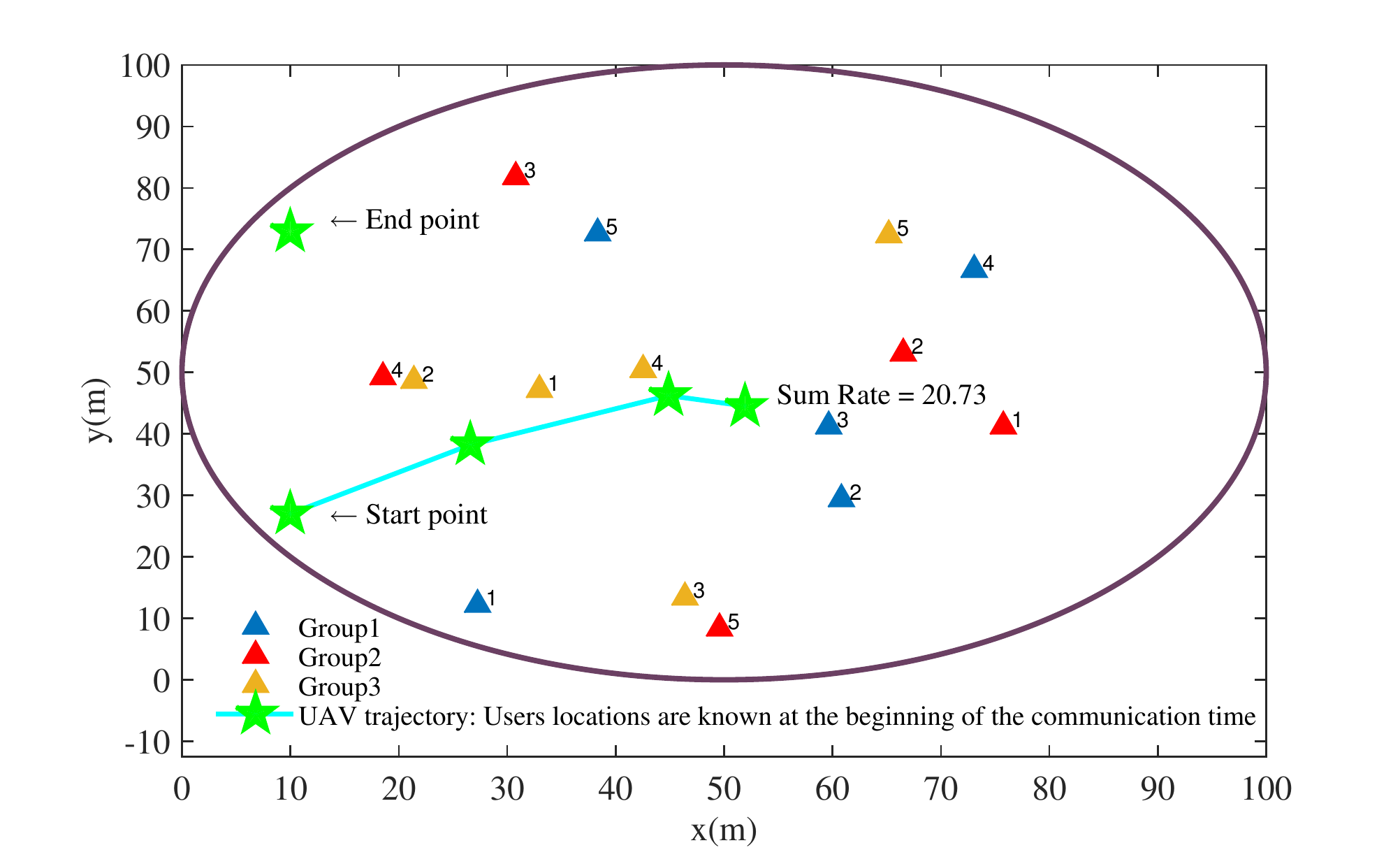}\label{mobileuser7d}}
	\subfigure[Users positions and UAV trajectory at the beggining time slot 5
	]
	{\includegraphics[width=0.495\linewidth]{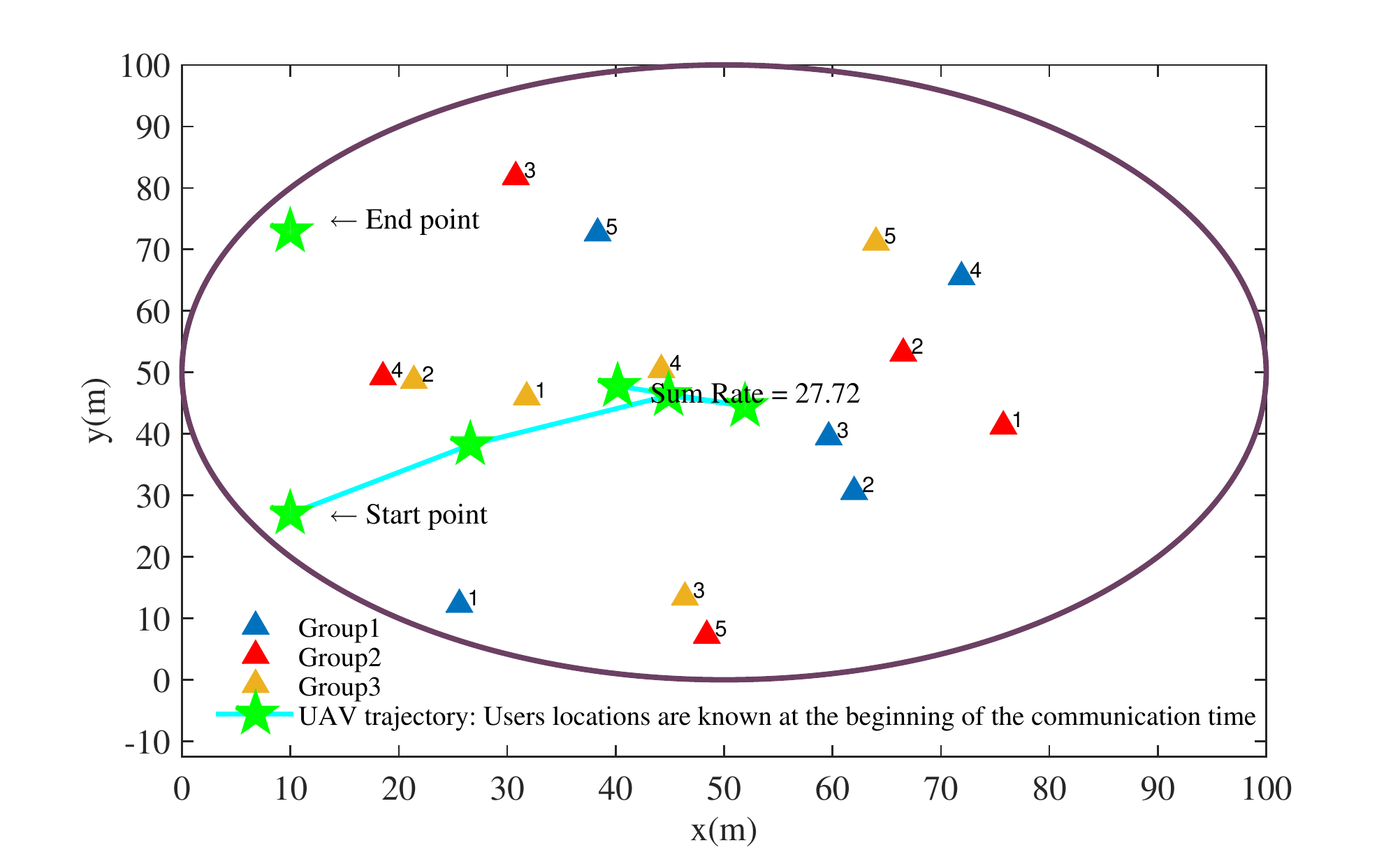}\label{mobileuser7e}}
	\subfigure[Users positions and UAV trajectory at the beggining time slot 6
	]
	{\includegraphics[width=0.495\linewidth]{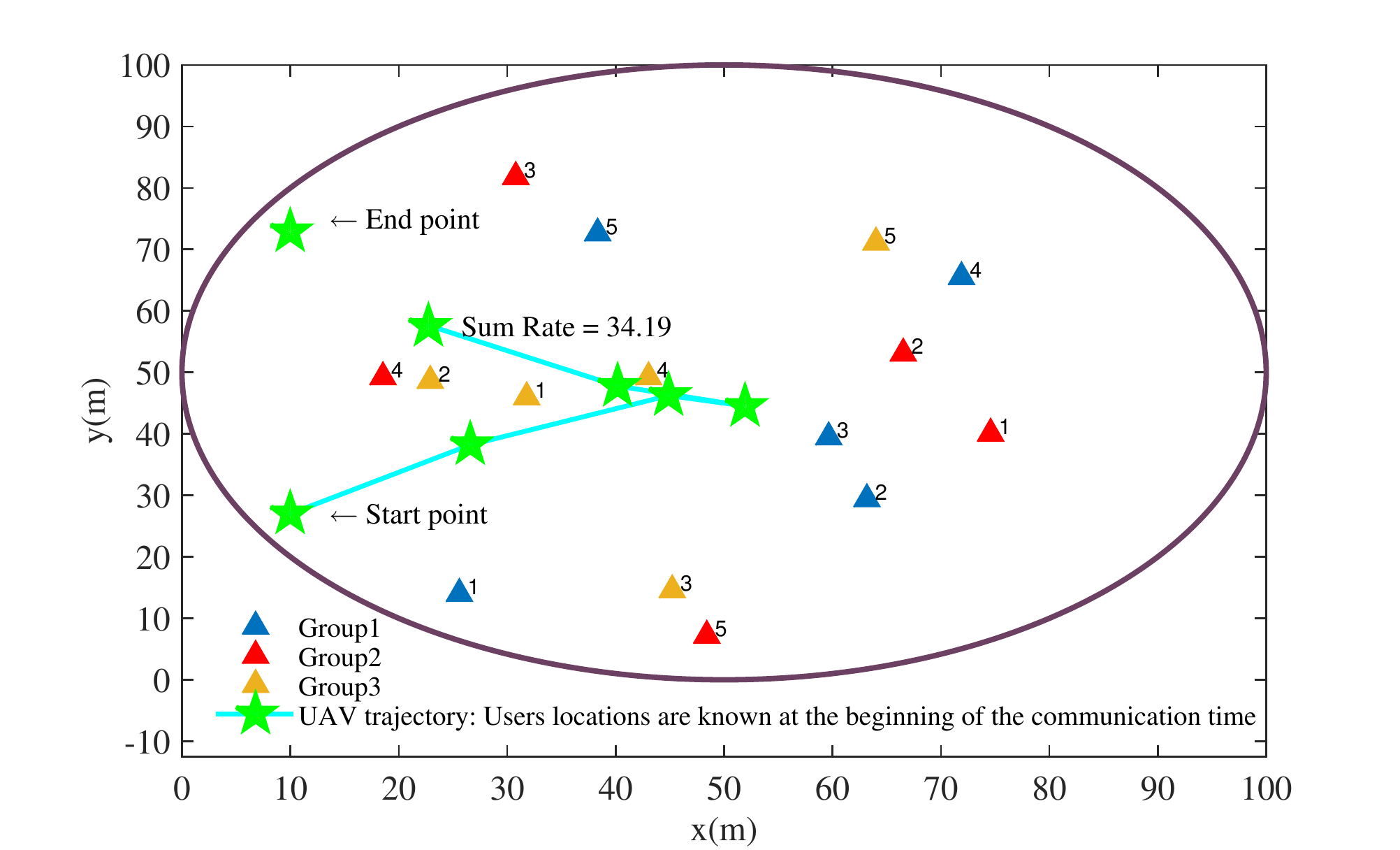}\label{mobileuser7f}}
	\subfigure[Final trajectory at the end of communication time
	]
	{\includegraphics[width=0.495\linewidth]{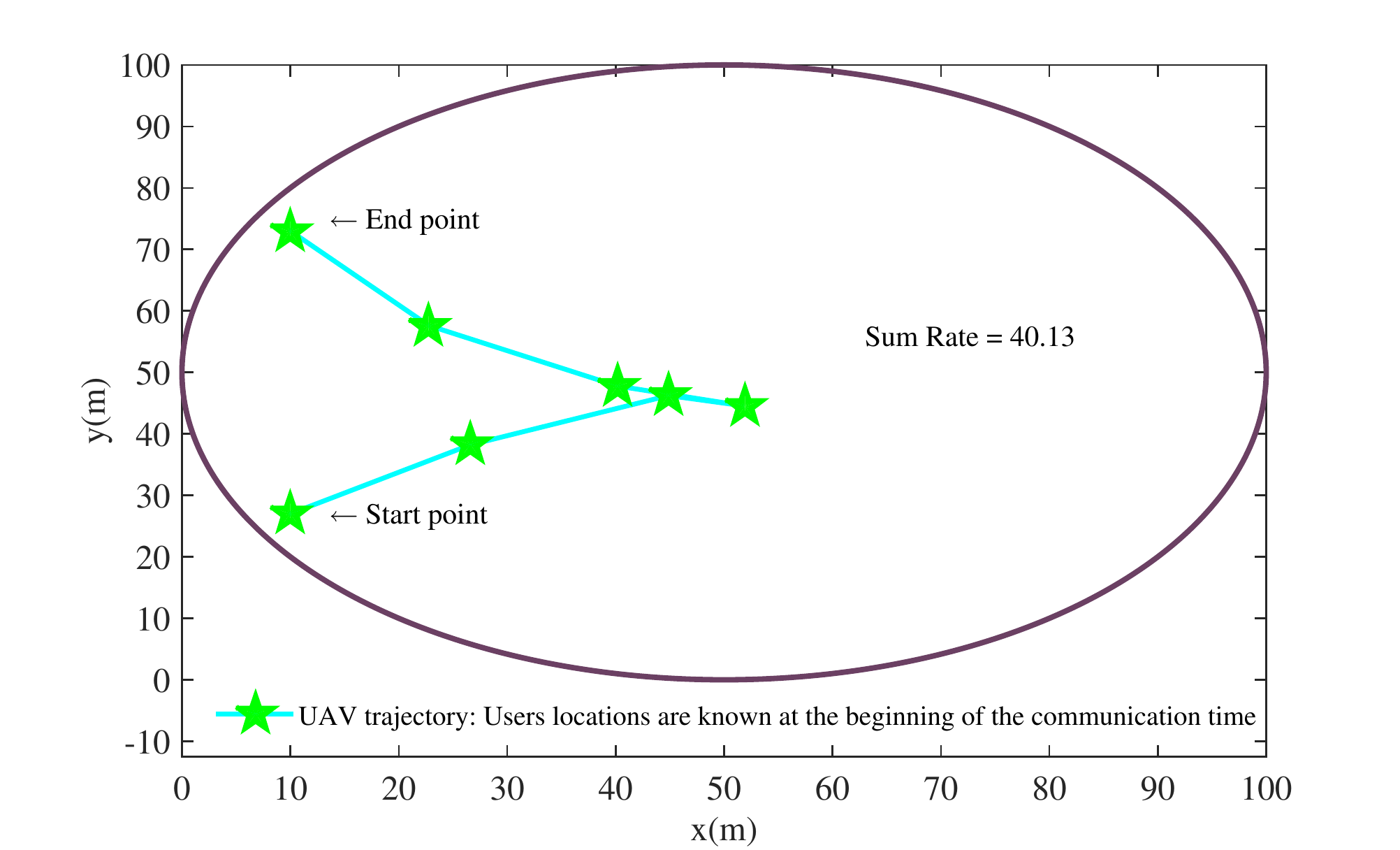}\label{mobileuser7g}}
	\caption{The trajectory of the UAV over time and the position of the UAV at the beginning  of each time slot for case of known location of users. System parameters are $G=3$, $T=12$s, $N=6$, $C_g^{\text{rsv}}[n]=0.25$ bps/Hz, $U_g=5, ~\forall g\in\mathcal{G}$.}
	\vspace*{4pt}\label{mobileuser}
\end{figure*}

		\textcolor{black}{
			With the approach of changing the number of users per group in each time slot, the last simulation is proposed. In all simulation parts, numbers of users are fixed but in the last, we show the variation of the number of users per group in each time slot in online scenario based on the Poisson distribution with parameter $\lambda$, that with the best study of us, none of the similar works didn't investigate it. As a realization, we show the variation of the number of users in time slots 1 and 2 i.e., in Fig. \ref{fig:13a} and Fig. \ref{fig:13b}. As you can see, Fig. \ref{fig:13a} show the number of users per group 1, 2, and 3 in time slot 1, which are 9, 7, and 5, respectively.  Fig. \ref{fig:13b} show the number of users per group 1, 2, and 3 in time slot 2, which are 9, 9, and 7, respectively, too.
			In Fig. \ref{fig:13c}, we evaluate the effect of power on the total rate with the help of different values of $\lambda$. Different values of $\lambda$ cause generate users with a different rate, therefore increasing $\lambda$ leads to increasing the number of users and if the added users have worse channel gain compared to existing ones, the total rate decreases or otherwise the total rate remains fixed. In Fig. \ref{fig:13c}, we run several times communication times, then we put the average of them here, for different values of $\lambda$.}

\section{Conclusion}\label{sec:conclusion}
In this paper we proposed a novel scheme for resource allocation in UAV assisted wireless network where the information is transmitted towards the network users by multicasting. We assumed that the UAV flies over a communication path \textcolor{black}{for a period of time over a communication path, and according to the concept of PD-NOMA, UAV sends the proportional message towards the network users of each multicast group, simultaneously}.
We aimed at jointly finding the UAV trajectory and the transmit power over the trajectory. 
\textcolor{black}{We considered offline mode from the perspective of; fixed users and mobile users with predictable location and an online mode that this case is closer to reality and more practical. In online mode, users are mobile and change their locations toward the communication time. Also, we investigated the variation of the number of users in each group and in each time slot for mobile users in an online scenario, too.} 
The proposed schemes are studied numerically for different values of the network parameters whose performances are confirmed by the results.

\ifCLASSOPTIONcaptionsoff
  \newpage
\fi

\bibliography{uav}	
\bibliographystyle{ieeetran}
%




\end{document}